\documentclass{ws-ijmpa}

\usepackage[super,compress]{cite}
\usepackage{wrapfig}

\newcommand{\als}{\alpha_s}
\newcommand{\ber}{\begin{eqnarray}}
\newcommand{\eer}{\end{eqnarray}}
\newcommand{\beq}{\begin{equation}}
\newcommand{\eeq}{\end{equation}}
\newcommand{\nn}{\nonumber}
\newcommand{\sgh}{\sigma(\omega,\vec{p})}
\newcommand{\href}{\underline}

\begin{document}

\markboth{\'Agnes M\'ocsy, P\'eter Petreczky, and Michael Strickland}
{Quarkonia in the Quark Gluon Plasma}

\catchline{}{}{}{}{}

\title{Quarkonia in the Quark Gluon Plasma}

\author{\'Agnes M\'ocsy}
\address{Department of Mathematics and Science, Pratt Institute, Brooklyn, NY 11205, USA\\
Physics Department, Brookhaven National Laboratory, Upton NY 11973 USA}

\author{P\'eter Petreczky}
\address{Physics Department, Brookhaven National Laboratory, Upton NY 11973 USA}

\author{Michael Strickland}
\address{Department of Physics, Kent State University, Kent OH 44242 USA }

\maketitle


\begin{abstract}
In this paper we review recent progress towards understanding the nature of quarkonia in the quark-gluon plasma.
We review the theory necessary to understand the melting of bound states due to color-screening,
including lattice results for the heavy quark potential, lattice results on the correlation functions related to the relevant spectral functions,
and the emergence of a complex-valued potential in high-temperature quantum chromodynamics. 
We close with a brief survey of phenomenological models of quarkonium suppression in relativistic heavy ion collisions.
\end{abstract}

\ccode{PACS numbers: 11.15Bt, 04.25.Nx, 11.10Wx, 12.38Mh}


\section{Introduction} 

The theory of strong interactions describes the fundamental force responsible for the stability of nuclei and the particles that themselves constitute nuclear matter, e.g. the proton, neutron, etc.  The theory describes the interaction and properties of the most elementary constituents of matter currently known:  quarks and gluons.  The quantum theory which describes the interaction of color-charge carrying quarks and gluons is called Quantum Chromodynamics (QCD).  QCD predicts that when the temperature of nuclear matter is increased above a certain threshold, strongly interacting matter undergoes a phase transition to a ``new'' state of matter referred to as the quark-gluon plasma (QGP).  Through this phase transition the degrees of freedom change from color-neutral hadrons to color-charged partons which are no longer confined to exist only inside color-neutral hadrons.  We note that calling this phase of matter ``new'' QGP is somewhat of a misnomer, since it is thought to be one of the oldest forms of matter in the universe, presumably present a millionth of a second after the birth of the universe in the Big Bang. QCD predicts that in the high temperature deconfined QGP phase, the interactions between quarks is color-screened much like electric charge is screened in an electromagnetic plasma. In addition, the QGP is also characterized by the restoration of a fundamental QCD symmetry, which is broken at low temperatures, called chiral symmetry. 

Due to non-linearities present in QCD there is as of yet no comprehensive analytic understanding of the theory.  For difficult non-perturbative questions, of which there are many in QCD, practitioners instead utilize numerical calculations of the QCD partition function on discrete lattices called lattice QCD.  As a result there are several supercomputer centers around the world utilizing state-of-the-art computational resources in order to aid our understanding of fundamental questions in QCD. As mentioned, lattice QCD is a quite reliable tool particularly for addressing the non-perturbative aspects of QCD.  The phase transition itself is highly non-perturbative and, in the immediate vicinity of the transition, one cannot rely on weak-coupling perturbation theory techniques written solely in terms of either hadronic or quark and gluon degrees of freedom.  Many modern finite temperature lattice QCD results have achieved a high-degree of precision and the results obtained are consistent between different groups using a variety of lattice discretization methods.  Unfortunately, the determination of a precise transition temperature is impossible since, as we will discuss below, the QCD phase transition is a smooth crossover and not a sharp first or second order phase transition.  

One can, however, see deconfinement of the partonic degrees of freedom through a rapid rise in thermodynamic quantities, e.g. energy density and pressure, around $T \sim $ 160 MeV, which indicates the liberation of quark and gluon degrees of freedom.  At temperatures of approximately $300~$MeV and above, the entropy density found in lattice studies has an approximately 20\% deviation from the entropy density of a non-interacting gas of quarks and gluons.  This deviation is well-described by resummed weak-coupling QCD calculations, indicating that at these temperatures partonic degrees of freedom in the form of quark and gluon quasiparticles are the relevant degrees of freedom.  At such temperatures it is clear that color-electric screening becomes relevant.  The transition from hadronic degrees of freedom to quark and gluon degrees of freedom is a smooth process occurring between 160 and 300 MeV and, therefore, as mentioned previously, there is no well-defined transition temperature. There is, however, a chiral symmetry restoration transition temperature that can be clearly identified.  Lattice studies find the chiral symmetry restoration temperature to be at  $T \sim 154(9)~$MeV. In the rest of this work, and in the majority of the literature, the chiral symmetry restoration temperature is referred to as the QCD critical temperature, $T_c$.    

\begin{wrapfigure}{r}{2in}
\begin{center}
\includegraphics[width=1.5in]{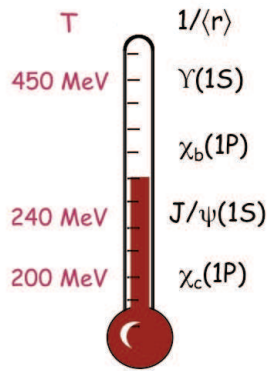} 
\end{center}
\caption{A cartoon of the QGP thermometer provided by the sequential melting of quarkonia taken from Ref.~\protect\refcite{Mocsy:2008eg}}
\label{fig:thermometer}
\end{wrapfigure}

Since the 1970's quarkonia, which are bound states of heavy flavor quarks (charm or bottom) and their antiquarks, have had an essential role in testing QCD, especially at zero temperature. Since the seminal work of Tetsuo Matsui and Helmut Satz in 1986\cite{Matsui:1986dk}, the behavior of quarkonia in high-temperature QCD has been considered a tell-tale signal for deconfinement and QGP production in heavy ion collisions. The original idea behind this proposal was that color screening in a deconfined plasma would reduce the binding of quarks, and thus prohibit the formation of heavy quark bound states. This ``melting'' of quarkonia in the QGP would manifest itself in the suppression of the quarkonia yields relative to the production without a QGP formation. Quarkonium melting at high temperature has been often quoted to be the QGP thermometer (see Fig.~\ref{fig:thermometer} for a cartoon rendition of the idea), serving as an essential tool to measure the plasma temperature. Various states with different masses have different sizes and thus are expected to be screened and dissociate in the QGP at different temperatures.  

As we discuss in detail in this review, the existence and the dissociation of quarkonia in the hot QGP medium is far from being this simple. The three leading complications are that (i) there are more than just color screening effects affecting the binding of a particular state in the QGP (there is for instance Landau damping), (ii) there exists a poorly understood non-perturbative intermediate temperature region just above the chiral critical transition (the aforementioned domain $160-300~$MeV) where identifying the relevant degrees of freedom, and thus their relevance for quarkonia must be carefully considered, and (iii) there exist purely nuclear initial state effects, collectively called cold nuclear matter effects (e.g. nuclear absorption, parton energy loss, and modifications of the parton distribution function), which must also be accounted for.

During the last two decades, quarkonia in a finite temperature medium have been studied using lattice QCD, but the results, more precisely the interpretation of the results, have not been without controversy. Effects of the hot medium are imprinted on the spectral function, the quantity which contains all the information in a given heavy quark-antiquark channel. Therefore, in principle, determining spectral functions at different temperatures should give us insight into the effects of a QGP on the bound states. For instance, the disappearance of a well-defined peak into the background of continuum states would be a clear indication for the melting of a state. This allows us to define what is referred to as a dissociation temperature (as we discuss below, we would like to draw attention to inconsistencies in the literature in how dissociation temperatures are precisely defined).  In this review we discuss state-of-the-art lattice QCD results relevant for in-medium quarkonium properties, including screening and time- and space-like heavy meson correlation functions.  Both of these are directly connected to the spectral functions that we seek and enable a determination of the static quark-antiquark interaction potential. We also briefly review what we have learned about quarkonia from these results. 

The formation, behavior, and modification of the properties of the quarkonia has also been addressed in phenomenological models, commonly referred to as potential models. We would like to highlight though, that the semantics is somewhat misleading, because potential models need not rely on phenomenological constructions of the heavy quark interaction potential, as was done in early studies. The potential can be derived directly from the QCD Lagrangian. At zero temperature, this was accomplished some time ago already. In recent years, finite temperature field theory derivations of the real part and, the previously ignored, imaginary part of the heavy quark potential have been completed.  We will refer to these QCD-derived approaches as potential models even, and discuss them in more detail below.  There are currently limitations to what questions such potential results can answer, making other phenomenological models necessary. Therefore, when educated guesses for the potential (based on high temperature QCD, and/or lattice QCD results for the free energy) are used to analyze quarkonia at finite temperature, or dynamical modeling is utilized to bridge from static spectral functions to their dynamical evolution in a cooling and expanding medium, we refer to these as phenomenological models.

In addition to the theoretical work, the QGP is studied in heavy-ion collision experiments at different collision energies at the SPS (CERN), RHIC (BNL) and the LHC (CERN). The QGP has been produced and found to behave as a nearly perfect fluid, although the detailed properties and behavior of this matter are still under scrutiny. These experimental programs dedicate significant effort to measuring quarkonium production. An important experimental observable is the nuclear modification factor, $R_{AA}$, which is the observed particle yield in nucleus-nucleus collisions (AA) relative to scaled proton-proton collisions (pp) where QGP formation is not expected. Quarkonium suppression has been observed in all of the above experiments. The ground state charmonium $J/\psi$ suppression was a key signature of deconfinement already at the SPS.  More recent experiments have led to the understanding that not all of the suppression is caused by the QGP: quarkonium suppression observed in proton-nucleus collisions (pA) indicates suppression due to cold nuclear matter effects. Understanding the wealth of experimental results requires understanding the medium modification of quarkonia and disentangling hot- and cold-matter effects. In view of the wealth of data available from RHIC and LHC on charmonium and bottomonium suppression, with particular emphasis on their ground states, the $J/\psi$ and the $\Upsilon$, we present our current theoretical understanding of quarkonia.

We note in closing that recently the question of quarkonia at finite temperature has been addressed using string theory-based AdS/CFT methods.  Due to the nascent nature of these studies, we feel it is still premature to draw firm conclusions from these studies concerning our understanding QCD and, in particular, heavy ion collision data.  Since space for this review is limited and these studies fall outside the immediate area of expertise of all three authors, we don't discuss them further in this review; however, we are encouraged that there may be an additional non-perturbative method for addressing the behavior of quarkonia in the quark gluon plasma.

Our review has the following structure: In Section 2 we present the state-of-the-art in theoretical methods for addressing questions concerning quarkonia in the QGP. 
In particular, we discuss relevant results from lattice QCD and from the effective field theory approach to potential models. 
In Section 3 we address quarkonium phenomenology, including quarkonia in a viscous QGP, the status of dynamical models,  sequential suppression, and cold nuclear matter effects. In Section 4 we present our conclusions and outlook for future studies.   

     
\section{Theoretical Methods}

\subsection{QCD at Non-zero Temperature}

The properties of hot strongly-interacting matter can be studied using first principles field
theoretical methods.
The QCD partition function can be written as a path integral over the gluon and quark fields
living in Euclidean space-time with a finite time direction $\tau \le 1/T$,
with $T$ being the temperature \cite{LeBellac:375551}.
The effects of the medium are encoded in the boundary conditions on the gluon and quark
fields: gluon fields obey periodic boundary conditions, while quark fields obey anti-periodic
boundary conditions.
This formalism is commonly known as the imaginary time formalism. Dynamical properties
of hot strongly interacting matter can be obtained from time-dependent correlation functions.
In the imaginary time formalism one can calculate correlation functions along the imaginary time axis.
These can be related to real time correlation functions by analytic continuation. However, in
many cases it is simpler to formulate the problem directly in real time. This can be done using
the real time formulation of thermal QCD, in which the time runs in the complex plane, first
along the real axis then coming back from infinity to $-i/T$ \cite{LeBellac:375551}.

The imaginary time formalism allows one to calculate the path integral numerically using
lattice discretization (see Refs.~\refcite{Petreczky:2012rq} and \refcite{Philipsen:2012nu} for recent reviews).
Unfortunately, lattice methods cannot be used in the real time
formulation and because of this the study of quarkonium properties in hot QCD medium is
a challenging problem. Still, our current understanding of QCD thermodynamics is largely
based on lattice results. In particular, our knowledge of the QCD equation of state, the
nature of the transition and the corresponding transition temperature, and the response of
the QCD medium to external static chromo-electric fields, all come from lattice QCD calculations.
Most lattice calculations have been performed using so-called improved staggered fermion
formulation. Variants of these formulations go under the acronyms asqtad, HISQ, p4, and stout
(see Ref. \refcite{Petreczky:2012rq} for a detailed discussion).

As already mentioned in the introduction, hot strongly interacting matter undergoes a phase transition 
to a QGP, in which the dominant degrees of freedom are partonic.  
A rapid rise in the thermodynamic quantities (pressure and energy density),  indicative of liberation of many new degrees of freedom, occurs around temperatures of $160~$MeV and can be associated with deconfinement. 
As mentioned in the introduction, at temperatures of about $300~$MeV the entropy density is only $10-20\%$ smaller than the entropy density
of non-interacting gas of quarks and gluons and these deviations from the ideal gas can be accounted for
by using a resummed weak coupling expansion \cite{Andersen:2009tc,Andersen:2010ct,Andersen:2011sf,Andersen:2011ug,Haque:2012my,Andersen:2012wr,Petreczky:2009at}.
The transition in QCD for physical light (u, d) and strange quark masses is found to be a crossover \cite{Aoki:2006we}.
Therefore, the definition of a transition temperature is non-trivial. 
As we mentioned in the introduction, the QGP
is characterized by chiral symmetry restoration (broken in the vacuum and at low $T$). The chiral aspect of the QCD transition
allows for a definition of a transition temperature since chiral symmetry is a good symmetry. The light quark masses are small compared to typical QCD scales, and therefore universality arguments allow one to
relate the crossover temperature to the phase transition temperature in the limit of zero light
quark masses \cite{Ejiri:2009ac,Kaczmarek:2011zz,Bazavov:2011nk}. It turns out that the physical light quark masses
are sufficiently small to allow such a definition and one obtains $T_c=154(9)$ MeV \cite{Bazavov:2011nk}. This value
is close to the inflection point of the renormalized chiral condensate \cite{Borsanyi:2010bp}. 
However, chiral symmetry in the strange quark sector is restored at significantly higher temperatures.
In addition to the equation of state, fluctuations of conserved charges have also been calculated in lattice QCD.
At low temperatures, massive hadrons are the carriers of different conserved charges and, therefore, the fluctuations
of conserved charges are suppressed by the corresponding Boltzmann factors. In fact, at low temperatures,
fluctuations of conserved charges are well-described by a gas of hadrons and hadronic resonances 
\cite{Borsanyi:2011sw,Bazavov:2012jq}. Similarly to the  energy density and the entropy density, fluctuations 
of conserved charges show a rapid rise in a narrow temperature interval, and for $T>300~$MeV are also well 
described by weak coupling techniques \cite{Petreczky:2009at,Petreczky:2009cr,Borsanyi:2011sw}.

In the pure glue $SU(N)$ theory there is a symmetry associated with the deconfinement transition, called the center of
$Z(N)$ symmetry. The partition function of this theory is invariant under gauge transformations, which
leave the fields periodic up to a phase factor that is an element of $Z(N)$ group
(this is a subgroup of the $SU(N)$ gauge group called the center).
This symmetry is
spontaneously broken at high temperatures. Therefore, deconfinement is a true phase transition in $SU(N)$ gauge theories.
The corresponding phase transition temperature is $292(2)~$MeV (obtained through units of the string tension $\sigma$ is $T/\sqrt{\sigma}=0.629(3)$
assuming $\sqrt{\sigma}=465~$MeV).
The center symmetry is broken by the presence of dynamical quarks, and as we discuss in the next section, for physical values of the 
light quark masses it is not a good symmetry. Therefore, for QCD a deconfinement transition temperature, that is similar to the deconfining transition in pure gauge theory, cannot be 
defined.

\subsection{Color Screening at High Temperatures}
\label{sec:screening}

The QGP
is characterized by color
screening. 
At high temperatures, color electric screening can be understood in terms of the low-momentum behavior of
the temporal component of the gluon propagator $D_{00}(k_0,{\bf k})$. The corresponding self-energy, or
the electric polarization function, in this limit is $\Pi_{00}(0,{\bf k} \rightarrow 0)=m_D$ with $m_D=N/3 g T\sqrt{1+N_f/(2 N)}~$, $m_D$ being the leading order Debye mass \cite{LeBellac:375551}.
Here $N_f$ is the number of light quark flavors: $N_f=0$
for pure gauge theory and $N_f=3$ for 2+1 flavor QCD (the effects of the strange quark mass are typically
very small at high temperatures). The above result is gauge independent and, up to color and quark flavor factors, 
it is identical to the QED result, $m_D=e T/3$ \cite{LeBellac:375551}. The transverse part of the static gluon propagator
is not screened in perturbation theory, i.e. $\Pi_T(0,{\bf k} \rightarrow 0)=0~$. This, however,
is not the case non-perturbatively, as we discuss this later.

To study color electric screening non-perturbatively one uses Polyakov loops.
As discussed in the previous subsection, the deconfining transition in $SU(N)$
gauge theories is a true phase transition related to $Z(N)$ symmetry.
The order parameters of this phase transition are the expectation value of the Polyakov loop and 
the Polyakov loop correlator
\begin{eqnarray}
L(T)&=&\langle \frac1N {\rm Tr} W(\vec{x}) \rangle, ~W(\vec{x})=\prod_{\tau=0}^{N_{\tau}-1} U_0(\tau,\vec{x}),\\
C_{PL}(r,T)&=&\frac{1}{N^2} \langle {\rm Tr} W(r) {\rm Tr} W(0) \rangle.
\end{eqnarray}
The Polyakov loop transforms non-trivially under the $Z(N)$ transformation. 
The expectation value of the Polyakov loop is zero in the confined phase and non-zero above
the transition temperature. The qualitative change in the behavior of the Polyakov loop and its correlator above the phase transition
temperature is related to color screening. 
The correlation function
of the Polyakov loop is related to the free energy of a static quark anti-quark pair \cite{McLerran:1981pb}.
But, as pointed out already in Ref.~\refcite{McLerran:1981pb}, the Polyakov loop
and the Polyakov loop correlator require renormalization to be interpreted as the free energy of an isolated static quark
or the free energy of a static quark anti-quark $(Q \bar Q)$ pair. More precisely, these quantities are related to the free energy 
difference of a system with static $Q\bar Q$ pair at some temperature and the system without $Q\bar Q$ pair at
the same temperature. 
Since in the zero temperature limit the free energy of a static
quark anti-quark pair should coincide with the static potential, the Polyakov loop renormalization is determined
by the normalization constant of the static potential, namely
\begin{eqnarray}
L_{ren}(T)&=&\exp(-c/(2 T)) L(T)=\exp(-F_{\infty}(T)/(2 T))\\
C_{PL}(r,T)&=&\exp(-F(r,T)/T+c/T),~~F_{\infty}(T)=\lim_{r \rightarrow \infty} F(r,T),
\end{eqnarray}
where $c$ is an additive normalization that ensures that the static potential has a certain value at a given
distance. In the above expressions we made it explicit that the free energy of an isolated static quark, $F_Q~$, is one-half of the free energy of an infinitely-separated $Q\bar Q$ pair. 
In the confined phase, the free energy of a static quark anti-quark pair is proportional to $\sigma(T) r$ at
large distances $r$, as expected in QCD. The effective temperature-dependent string tension $\sigma(T)$
is non-zero below the phase transition temperature \cite{Kaczmarek:1999mm,Digal:2003jc}. 
Consequently, the free energy of an isolated static quark is infinite and 
$L_{ren}(T)=0$. In the deconfined phase, the free energy of an infinitely separated $Q\bar Q$ pair, $F_{\infty}(T)~$, is finite due to color screening. In particular, at
high temperatures $F_{\infty}=-(N^2-1) \alpha_s m_D/(2 N)$ at leading order. This result follows
from the infrared behavior of the self energy $\Pi_{00}(0,k)$ discussed above.
The next-to-leading order correction for $F_{\infty}$ has
been calculated in Refs.~\refcite{Burnier:2009bk} and \refcite{Brambilla:2010xn} and was found to be small. 
For the free energy of a static $Q\bar Q$ pair at distance $r$ we have at leading order
\cite{Nadkarni:1986cz}
\begin{equation}
F(r,T)=-\frac{1}{N^2} \frac{\alpha_s^2}{r^2} \exp(-2 m_D r)-\frac{N^2-1}{2N} \alpha_s m_D. 
\end{equation}
Note, that this result is in contrast to the free energy of static charges in QED which is $-\alpha \exp(-m_D r)/r-\alpha m_D$,
and is the consequence of non-Abelian nature of interactions. A static quark anti-quark pair could be either
in a singlet or in an octet state, and thus (in a fixed gauge) one can define the so-called color singlet and
octet free energy \cite{Nadkarni:1986cz,Nadkarni:1986as} \footnote{The terms color singlet free energy and octet free energy are misleading as only
$F(r,T)$ has the meaning of the free energy of static quark anti-quark pair. Furthermore, at low temperatures both $F_1$ and
$F_8$ are determined by color singlet asymptotic states \cite{Jahn:2004qr}. The color screening of the static quark quark
interaction was discussed in Ref.~\refcite{Doring:2007uh}, while color screening of charges in higher representations was studied in Refs.~\refcite{Gupta:2007ax} and \refcite{Mykkanen:2012ri}.}
\begin{eqnarray}
\displaystyle
\exp(-F_1(r,T)/T+c/T)&=&
\frac{1}{N} \langle {\rm Tr}[W^{\dagger}(x) W(y)]\rangle,\label{F1def}\\
\displaystyle
\exp(-F_8(r,T)/T+c/T)&=&
\frac{1}{N^2-1}
\langle {\rm Tr} W^{\dagger}(x)  {\rm Tr} W(y) \rangle \nonumber\\
\displaystyle
&-&\frac{1}{N (N^2-1)} \langle {\rm Tr}\left[W^{\dagger}(x)  W(y)\right]\rangle.\label{Fadef}
\end{eqnarray}
One can view the above correlators as correlation functions of singlet and octet
static meson operators of size $r$ in Coulomb gauge evaluated at $\tau=1/T$ \cite{Jahn:2004qr,Bazavov:2008rw}.
Then, the Polyakov loop correlator can be written as the thermal average over 
the singlet and the octet contributions,  
\cite{McLerran:1981pb,Nadkarni:1986cz,Nadkarni:1986as}
\begin{equation}
C_{PL}(r,T)=\frac{1}{N^2} \exp(-F_1(r,T)/T+c/T)+\frac{N^2-1}{N^2} \exp(-F_8(r,T)/T+c/T).
\label{decomp}
\end{equation}
The singlet and octet free energies can be calculated at high temperature in
leading order hard-thermal-loop (HTL) approximation \cite{Petreczky:2005bd}, resulting in
\begin{eqnarray}
\displaystyle
F_1(r,T)&=&-\frac{N^2-1}{2N} \frac{\alpha_s}{r} \exp(-m_D r)-\frac{(N^2-1) \alpha_s m_D }{2 N},
\label{f1p}\\
\displaystyle
F_8(r,T)&=&\frac{1}{2N} \frac{\alpha_s}{r} \exp(-m_D r)-\frac{(N^2-1) \alpha_s m_D}{2 N}.
\label{f3p}
\end{eqnarray}
The singlet and octet free energies are gauge independent at this order. 
Inserting the above expressions into Eq. (\ref{decomp}) and expanding
in $\alpha_s~$, one can see that the leading order contributions from the singlet and
octet channels cancel each other and one recovers the leading order
result for $F(r,T)$ in Eq. (\ref{decomp}). 
Thus the non-Coulombic nature  of the $Q \bar Q$ free energy is due to the partial 
cancellation of the singlet and octet contributions at leading order.
For a long time it was not clear how to generalize
the decomposition of the free energy into the singlet and octet contributions beyond
leading order. Recently, using the effective theory approach that will be discussed below,
it was shown that this decomposition indeed holds at short distances \cite{Brambilla:2010xn}. 
As we will see below, the singlet free energy turns out to be rather 
useful when studying screening numerically in lattice calculations.

\begin{figure}
\includegraphics[width=6.1cm]{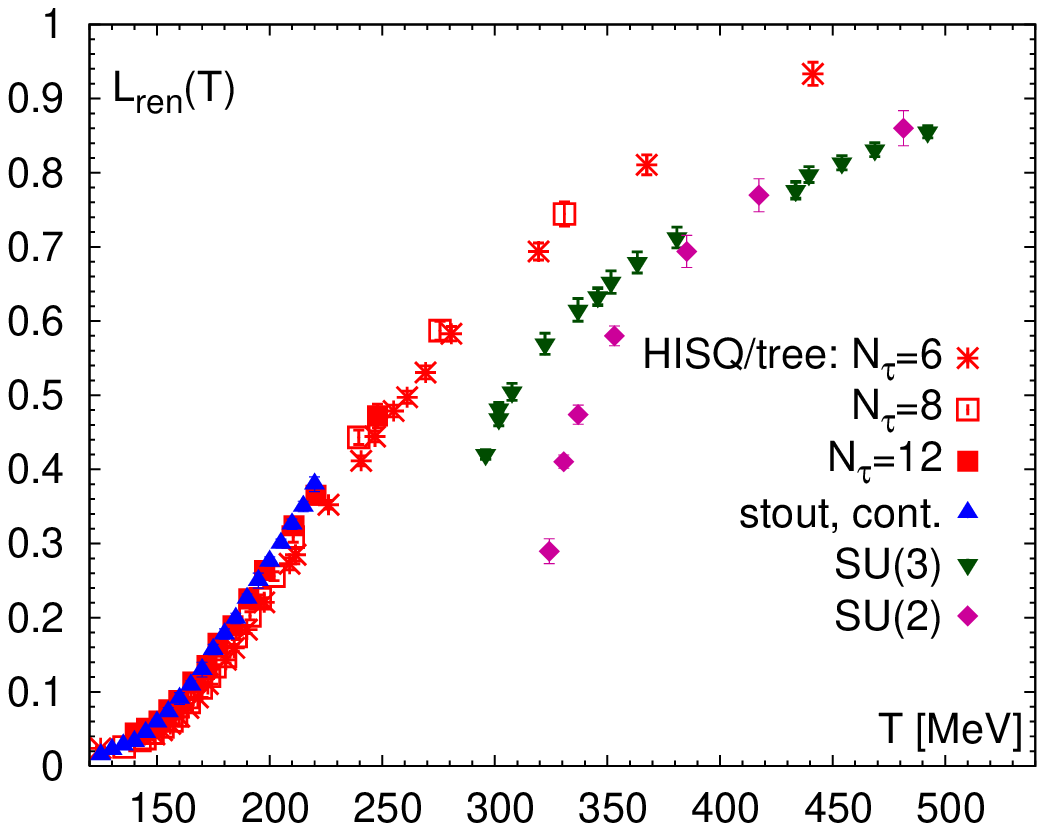}
\includegraphics[width=6.4cm]{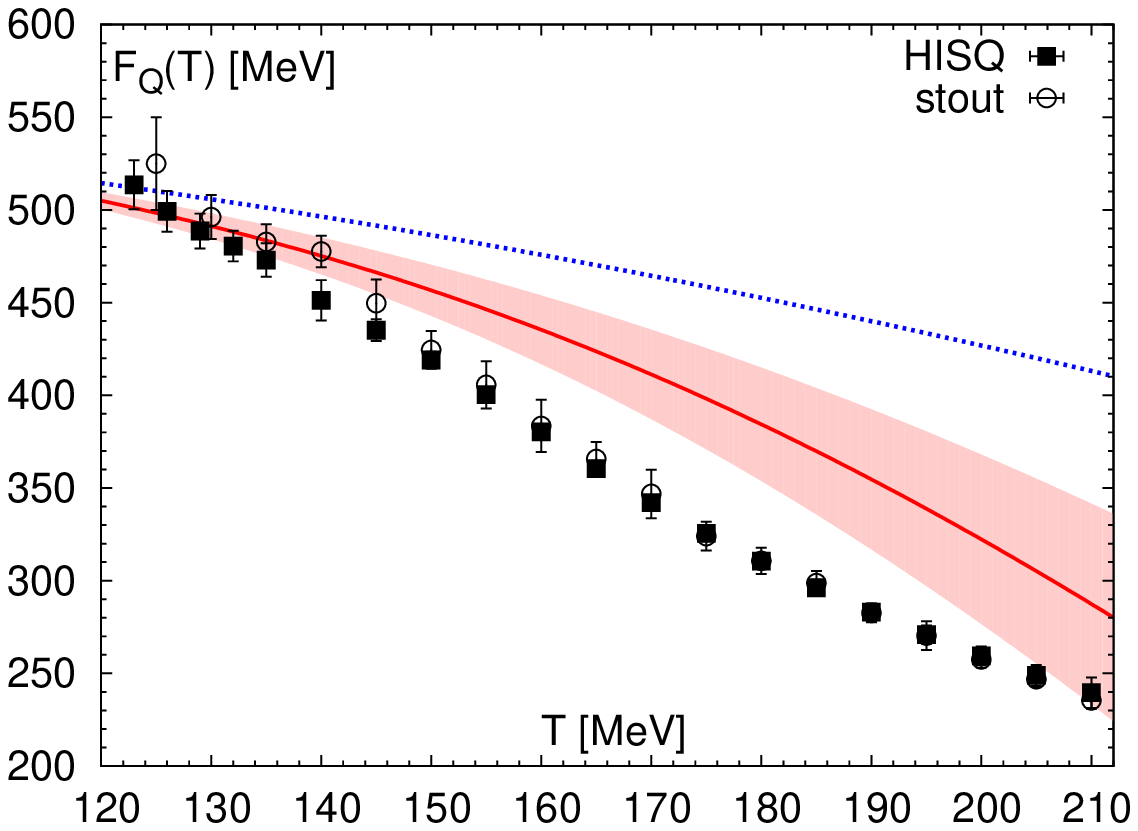}
\caption{The Polyakov loop as function of temperature in 2+1 flavor QCD \cite{Borsanyi:2010bp,Bazavov:2011nk}
and in pure gauge theory \cite{Digal:2003jc,Kaczmarek:2002mc}.
The right panel shows the free energy of a static quark $F_Q=F_{\infty}/2$ as function of temperature
and compared with the hadron resonance gas model of static-light hadrons \cite{Bazavov:2013yv}
}
\label{fig:lren}
\end{figure}

As already mentioned, the QCD partition function does not possesses $Z(3)$ symmetry. This symmetry is broken by the quark
contribution. In physical terms this means that a static quark anti-quark pair can be
screened already in the vacuum by light dynamical quarks. The free energy of a static $Q \bar Q$ pair does not
rise linearly with distance $r$ indefinitely but saturates at some distance, i.e. we see string breaking.
For light dynamical quarks the corresponding free energy, $F_{\infty}~$, at very low temperatures is twice the binding energy 
of static-light meson and thus is proportional to $\Lambda_{QCD}$. 
As the temperature increases excited static light mesons, as well as baryon states 
with one static quark will contribute to the expectation value of the Polyakov loop \cite{Megias:2012kb}.
Using the spectrum of known hadron states with a heavy quark or using the spectrum
of static light hadrons, the expectation value of the Polyakov loop can be calculated
in the approximation of a non-interacting gas of these hadrons  \cite{Megias:2012kb,Bazavov:2013yv}.
In Fig. \ref{fig:lren} we show the recent lattice results for the renormalized Polyakov loop
in 2+1 flavor QCD with physical quark masses extrapolated to the continuum limit \cite{Borsanyi:2010bp,Bazavov:2013yv} 
and compare these to the results in pure gluodynamics \cite{Digal:2003jc,Kaczmarek:2002mc}.
The temperature scale in pure gauge theory was set using the value $\sqrt{\sigma}=465~$MeV for
the zero temperature string tension. At high temperatures the behavior of the Polyakov loop
in pure glue theory and QCD look similar, at least qualitatively. The difference appears
when approaching the transition region. The Polyakov loop behaves smoothly across
the transition region. We also show the comparison of the free energy of a static quark, $F_Q=F_{\infty}/2~$, with the prediction of
a resonance gas of static-light states \cite{Bazavov:2013yv}, which seems to work well within
the uncertainties of the model up to temperatures of about $140~$MeV.

\begin{figure}
\includegraphics[width=5.9cm]{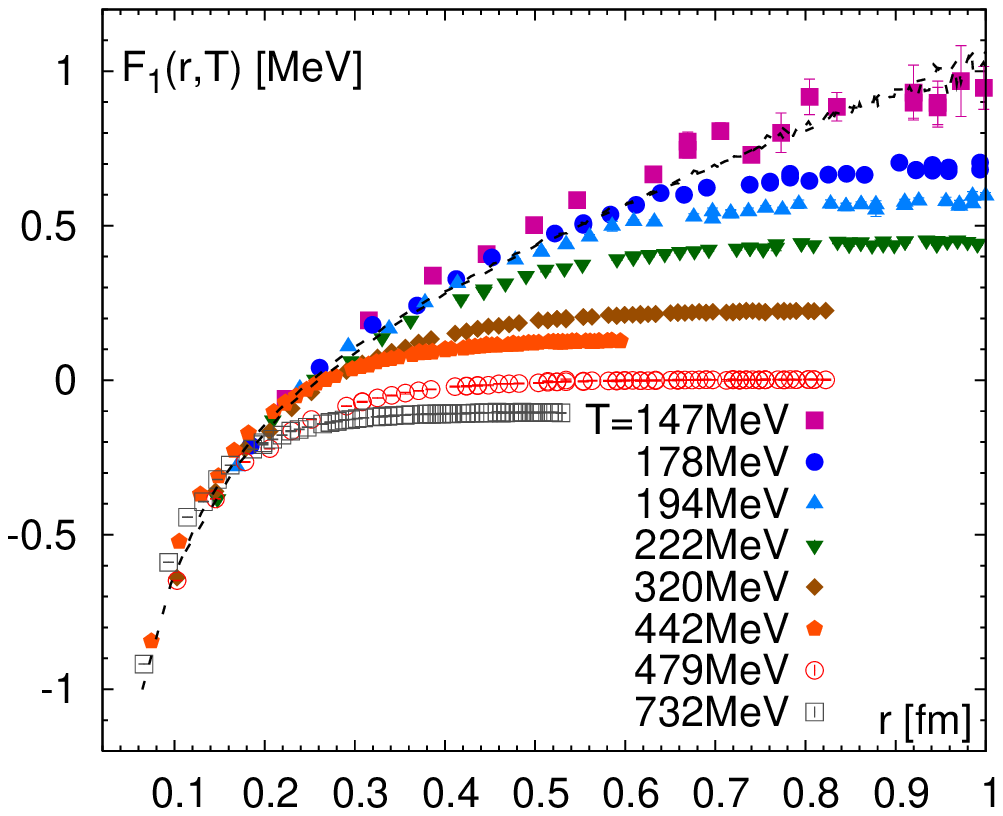}
\includegraphics[width=6.6cm]{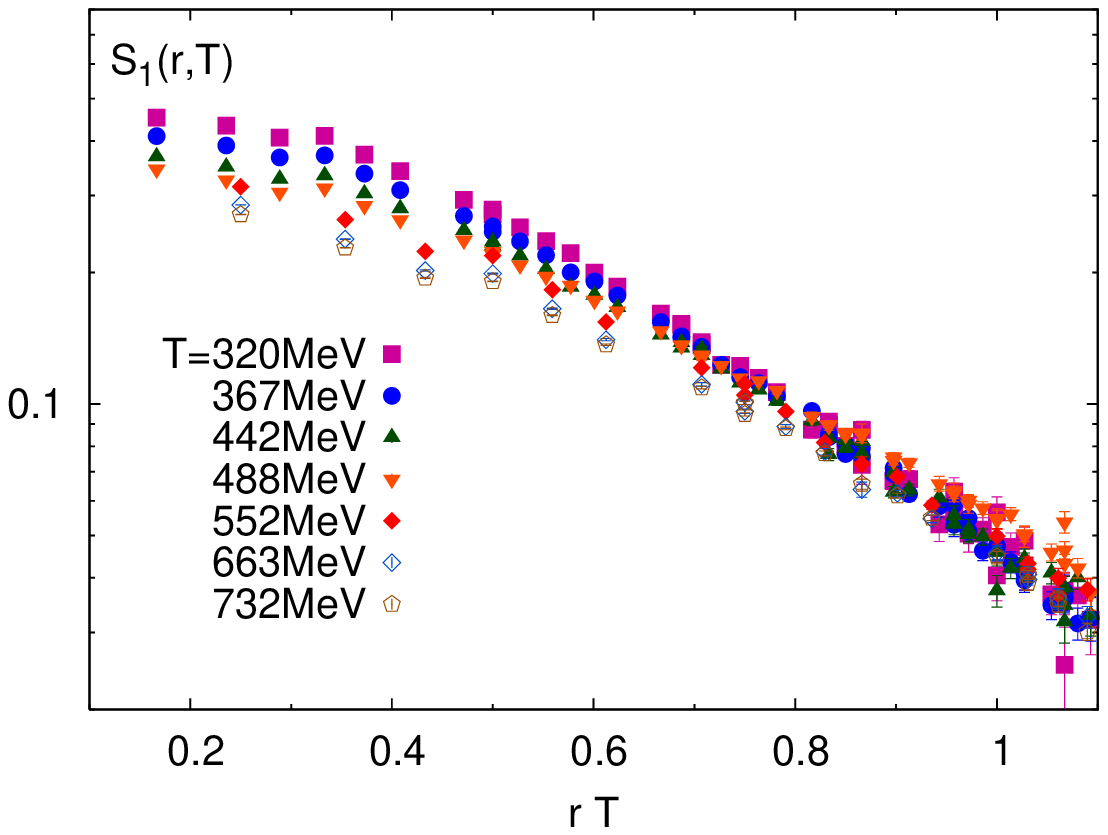}
\vspace*{-0.3cm}
\caption{The singlet free energy (left) and the screening function (right) as 
function of quark separation distance $r$ at different temperatures calculated with the HISQ action \cite{Bazavov:2012fk}.}
\label{fig:f1}
\vspace*{-0.3cm}
\end{figure}

The free energy of a $Q \bar Q$ pair, as well as the singlet free energy have been studied in 2+1 flavor
QCD using the HISQ action and $16^3 \times 4$ and $24^3 \times 6$ lattices \cite{Bazavov:2012fk}.
The numerical results for the singlet free energy are shown in Fig. \ref{fig:f1}. At short distances the
singlet free energy agrees with the zero temperature potential calculated in Ref.~\refcite{Bazavov:2011nk},
while at large distances it approaches a constant value $F_{\infty}(T)~$, equal to the excess free energy
of two isolated static quarks. As the temperature increases, the deviation from the zero-temperature potential
shows up at shorter and shorter distances as the consequence of color screening. 
For temperatures $T>200~$MeV there are large deviations from the zero temperature potential already
at distances $r$ that are similar to charmonium size. Thus we would expect strong modification and/or
dissolution of charmonium states. To explore the screening
behavior in Fig. \ref{fig:f1} we also show the combination $S(r,T)=r \cdot (F(r,T)-F_{\infty}(T))$ which
we call the screening function. The screening function 
should decay exponentially. We indeed observe the exponential decay of this quantity at distances larger
than $0.8/T~$. Thus at high temperatures the behavior of the singlet free energy expected from the weak-coupling 
calculations seems to be confirmed by lattice QCD, at least qualitatively.
Let us also mention that at high temperatures
the behavior of the singlet free energy is similar to that observed in pure gauge theory \cite{Digal:2003jc,Kaczmarek:2002mc}.

\begin{figure}
\includegraphics[width=5.9cm]{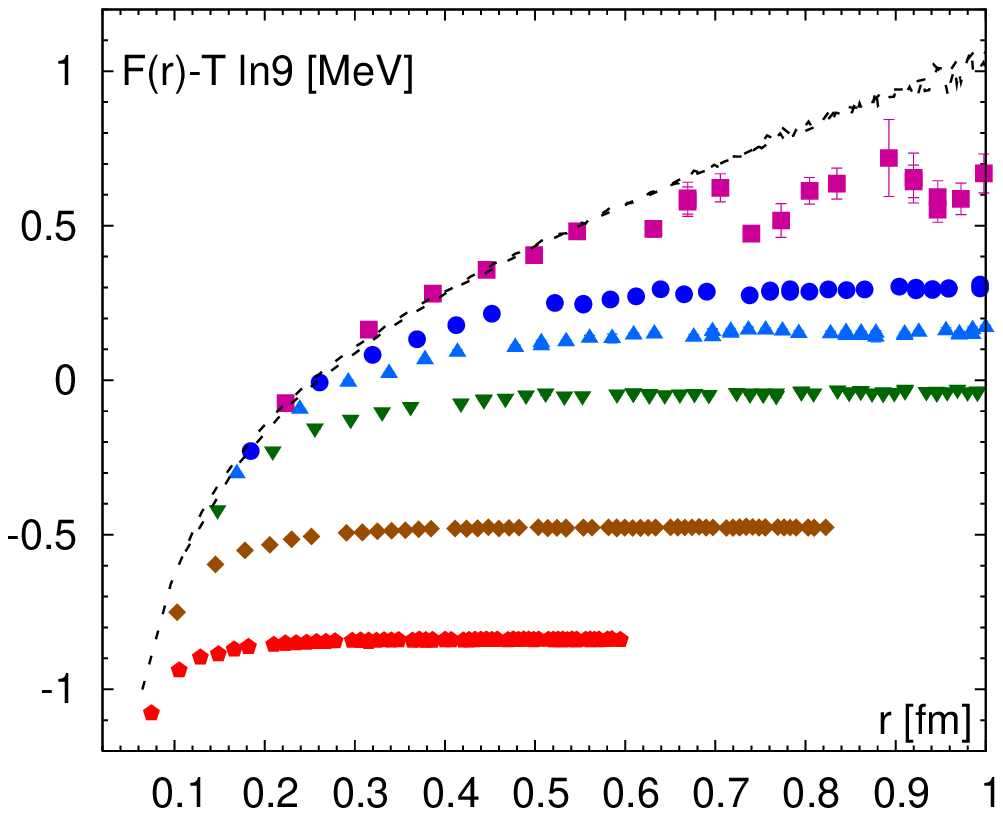}
\includegraphics[width=6.6cm]{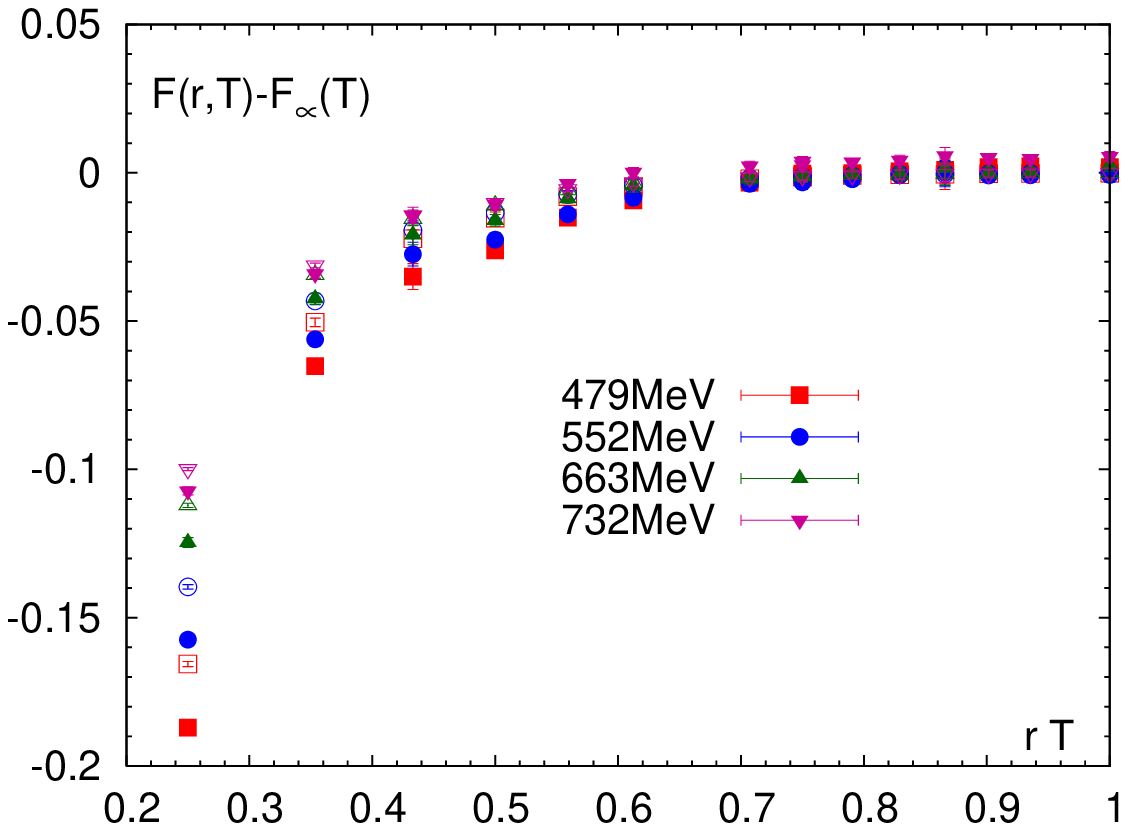}
\vspace*{-0.3cm}
\caption{The free energy of static $Q\bar Q$ pair (left) and the 
difference $F(r,T)-F_{\infty}(T)$ (right) calculated with HISQ action as 
function of quark separation distance $r$ at different temperatures. In the right panel the filled symbols correspond
to the lattice data, while the open symbols correspond to the values reconstructed from
the singlet free energy \cite{Bazavov:2012fk}.
The legend in the left panel is the same as in Fig. \ref{fig:f1} (left).
}
\vspace*{-0.3cm}
\label{fig:f}
\end{figure}

In Fig. \ref{fig:f} we show lattice results for the free energy of a static $Q \bar Q$ pair as function of the 
distance at different temperatures. At short distances and low temperatures the free energy is expected to be
dominated by the singlet contribution and we expect it to be equal to the zero temperature potential up to the
term $T \ln 9~$, term coming from the normalization (see the discussion in the previous section). 
Therefore, in the figure, the numerical results have been shifted by $-T \ln9~$.
Indeed, for the smallest temperature and the shortest distances $F(r,T)-T \ln 9$ is equal to the zero temperature potential
shown as the dashed black line. At higher temperatures $F(r,T)$ is very different from the zero-temperature potential.
At large distances the free energy approaches a constant value $F_{\infty}(T)$ that
decreases with increasing temperatures, confirming expectations (see discussions above). The temperature-dependence of $F(r,T)$ is
much larger than that of the singlet free energy. This is presumably due to the partial cancellation of the singlet and octet
contribution discussed above. To verify this assertion we calculated $F(r,T)-F_{\infty}(T)$ using the numerical data for 
$F_1(r,T)-F_{\infty}(T)$ and the leading order relation  $(F_1(r,T)-F_{\infty}(T))/(F_8(r,T)-F_{\infty}(T))=-8~$.
The corresponding results are shown in the right panel of Fig. \ref{fig:f}. As one can see from the figure,
the numerical data for $F(r,T)$ are in reasonable agreement with the ones reconstructed from this procedure. The
reconstruction works better with increasing temperature. Thus the expected cancellation of the singlet and octet
contributions to the free energy of static $Q\bar Q$ pair seems to be confirmed by lattice calculations.
The numerical results discussed above are similar to the one obtained with p4 action \cite{Kaczmarek:2007pb,Petreczky:2010yn} for $T>200~$MeV.

The qualitative features of the singlet free energy discussed above are not specific to the Coulomb gauge.
In fact, the singlet free energy defined in terms of Wilson loops shows very similar behavior to the
one calculated in the Coulomb gauge \cite{Bazavov:2008rw}. Furthermore, the Debye screening mass can also be defined from the long
distance behavior of the electric gluon propagator 
\cite{Heller:1995qc,Heller:1997nqa,Karsch:1998tx,Cucchieri:2001tw,Nakamura:2003pu,Cucchieri:2012nx}.
Calculations in $SU(2)$ and $SU(3)$ gauge theories 
show no gauge dependence of the extracted Debye mass within statistical errors \cite{Cucchieri:2001tw,Nakamura:2003pu}.
The extracted screening masses  are in agreement with the ones obtained from the singlet free energy.
Color magnetic screening can be also studied in this approach 
\cite{Heller:1995qc,Heller:1997nqa,Karsch:1998tx,Cucchieri:2001tw,Nakamura:2003pu,Cucchieri:2012nx}, though
the long distance behavior of the static transverse gluon propagator is not always exponential \cite{Cucchieri:2000cy,Cucchieri:2001tw}
and thus the corresponding screening mass cannot be defined.

\subsection{Quarkonium Spectral Functions and Euclidean Correlators on the Lattice}
\label{sec:spf}

In-medium meson properties are encoded in meson spectral functions. 
The spectral function $\sgh$ for a given 
meson channel $H$ in a system at temperature $T$ can be defined 
through the Fourier-transform of the real time two-point function
$D^{>}$ and $D^{<}$ of the meson current $J_H~$, or equivalently, as the imaginary part of 
the Fourier-transformed retarded 
correlation function \cite{LeBellac:375551},
\ber
\sgh &=& \frac{1}{2 \pi} (D^{>}_H(\omega, \vec{p})-D^{<}_H(\omega, \vec{p}))
\nonumber\\
&&
=\frac{1}{\pi} Im D^R_H(\omega, \vec{p}) \nonumber \\
 D^{>(<)}_H(\omega, \vec{p}) &=& \int_{-\infty}^{\infty} dt \int d^3 x 
 e^{i \omega t -i \vec{p} \cdot \vec{x}} D^{>(<)}_H(t,\vec{x}) \label{eq.defspect} \\
D^{>}_H(t,\vec{x}) &=& \langle
J_H(t, \vec{x}) J_H(0, \vec{0}) \rangle \nonumber\\
D^{<}_H(t,\vec{x}) &=& 
\langle J_H(0, \vec{0}) J_H(t,\vec{x}) \rangle , t>0 \
\eer
Here we study local meson operators of the form 
\beq
J_H(t,x)=\bar q(t,\vec{x}) \Gamma_H q(t,\vec{x})
\label{cont_current}
\eeq
where $q(t,\vec{x})$ is the quark field operator and
\beq
\Gamma_H=1,\gamma_5, \gamma_{\mu}, \gamma_5 \gamma_{\mu}, \gamma_{\mu} \gamma_{\nu}
\eeq
for scalar, pseudo-scalar, vector, axial-vector and tensor channels. 
The relation of these quantum number channels to different meson states is given
in Table \ref{tab.channels}. 

\begin{table}[t]
\tbl{Meson states in different channels for light, charm, and bottom quarks.
}
{
\begin{tabular}{|c|c|c||c|}
\hline
$\Gamma$ & $^{2S+1}L_{J}$ & $J^{PC}$ & $u\overline{u}$\\\hline
$\gamma_{5}$ & $^{1}S_{0}$ & $0^{-+}$ & $\pi$\\
$\gamma_{s}$ & $^{3}S_{1}$ & $1^{--}$ & $\rho$\\
$\gamma_{s}\gamma_{s^{\prime}}$ & $^{1}P_{1}$ & $1^{+-}$ & $b_{1}$\\
$1$ & $^{3}P_{0}$ & $0^{++}$ & $a_{0}$\\
$\gamma_{5}\gamma_{s}$ & $^{3}P_{1}$ & $1^{++}$ & $a_{1}$\\
\hline
\end{tabular}
\begin{tabular}{|cc|}\hline
$c\overline{c}(n=1)$ & $c\overline{c}(n=2)$\\\hline
$\eta_{c}$ & $\eta_{c}^{^{\prime}}$\\
$J/\psi$ & $\psi^{\prime}$\\
$h_{c}$ & \\
$\chi_{c0}$ & \\
$\chi_{c1}$ & \\
\hline
\end{tabular}
\begin{tabular}{|cc|}\hline
$b\overline{b}(n=1)$ & $b\overline{b}(n=2)$\\
\hline
$\eta_b$ & $\eta_b'$ \\
$\Upsilon(1S)$ & $\Upsilon(2S)$\\
$h_b$ & \\
$\chi_{b0}(1P)$& $\chi_{b0}(2P)$\\
$\chi_{b1}(1P)$& $\chi_{b1}(2P)$\\
\hline                                                                   
\end{tabular}
}
\label{tab.channels}
\end{table}

The correlators $D^{>(<)}_H(t,\vec{x})$ satisfy the well-known Kubo-Martin-Schwinger (KMS) condition \cite{LeBellac:375551}
\beq
D^{>}_H(t,\vec{x})=D^{<}_H(t+i/T,\vec{x}).
\label{kms}
\eeq
Inserting a complete set of
states and using Eq. (\ref{kms}), one obtains the expansion
\ber
&
\sgh = \frac{1}{Z} \sum_{m,n} e^{-E_n / T} \times \nonumber\\ 
&
|\langle n | J_H(0) | m \rangle|^2 \left(\delta^4(p_\mu + k^n_\mu - k^m_\mu)- 
\delta^4(p_\mu + k^m_\mu - k^n_\mu)\right)
\label{eq.specdef}
\eer
where $Z$ is the partition function, 
$k^{n(m)}$ refers to the four-momenta of the state $| n (m) \rangle $ and $p_{\mu}=(\omega,\vec{p})$.

A stable meson state contributes a $\delta$-function-like
peak to the spectral function:
\beq
\sgh = | \langle 0 | J_H | H \rangle |^2 \epsilon(\omega)
\delta(p^2 - M^2),
\label{eq.stable}
\eeq
where $M_H$ is the mass of the state and $\epsilon(p_0)$ is the sign function. For 
a quasi-particle in the medium, one obtains a smeared peak, with the width
being the  thermal width. 
As one increases the temperature, the width increases and at sufficiently high temperatures, the contribution from the meson state in the spectral function may be sufficiently broad, and thus it is no longer meaningful to speak of it as a well-defined state. 

In finite temperature lattice calculations, one calculates Euclidean time propagators, usually
projected to a given spatial momentum:
\beq
G(\tau, \vec{p}) = \int d^3x e^{i \vec{p}.\vec{x}} 
\langle J_H(\tau, \vec{x}) J_H(0,
\vec{0}) \rangle .
\eeq
This quantity is an analytical continuation
of $D^{>}_H(t,\vec{p})$
\beq
G(\tau,\vec{p})=D^{>}_H(-i\tau,\vec{p}).
\label{cont}
\eeq
Using this together with the periodicity of the time direction, 
we obtain the following integral representation for the Euclidean time correlator
\ber
G(\tau, \vec{p}) &=& \int_0^{\infty} d \omega
\sgh K(\omega, \tau), \label{eq.spect} \nn\\
K(\omega, \tau) &=& \frac{\cosh(\omega(\tau-1/2
T))}{\sinh(\omega/2 T)}.
\label{eq.kernel}
\eer
This equation is the basic equation for {\it extracting} the spectral
function from meson correlators. 
Equation (\ref{eq.kernel}) is valid in the continuum. 
Formally, the same spectral representation can be written for
the Euclidean correlator calculated on the lattice $G^{lat}(\tau,\vec{p})$.
The corresponding spectral function, however, will be distorted by the effect
of the finite lattice spacing, in particular, the spectral function is zero above 
certain energy $\omega>\omega_{max}$.
These distortions have been calculated in the
free theory \cite{Karsch:2003wy,Aarts:2005hg}. 

Early attempts to reconstruct  charmonium spectral functions from Euclidean 
quarkonium correlators in quenched approximation (pure glue theory) using
the Maximum Entropy Method (MEM) have
been presented in Refs.~\refcite{Umeda:2002vr,Asakawa:2003re}, and \refcite{Datta:2003ww}. The reconstructed spectral functions contained some peak-like structures that survived up to temperatures as high as $1.6\,T_c$ and even higher, to $2\,T_c$. These structures have been interpreted as
the charmonium ground state surviving into the deconfined phase. This is contrary
to the picture of color screening obtained from the singlet free energy.
Similar results have been presented in Ref.~\refcite{Aarts:2007pk} for two flavor QCD.
It turned out, however, that the above conclusion
was premature: As we discuss below, quarkonium correlators are not particularly sensitive to modifications in the spectral function due to the dissolution of the quarkonium bound states. Detailed systematic studies revealed that with more realistic priors in the MEM analysis there is no indication of the existence of the bound states \cite{Jakovac:2006sf,Ding:2012sp} (for the discussion of MEM systematics also see Ref.~\refcite{Wetzorke:2001dk}).

\subsubsection{Temperature-dependence of the Euclidean Correlators}

\begin{figure}
\centerline{\includegraphics[width=8cm]{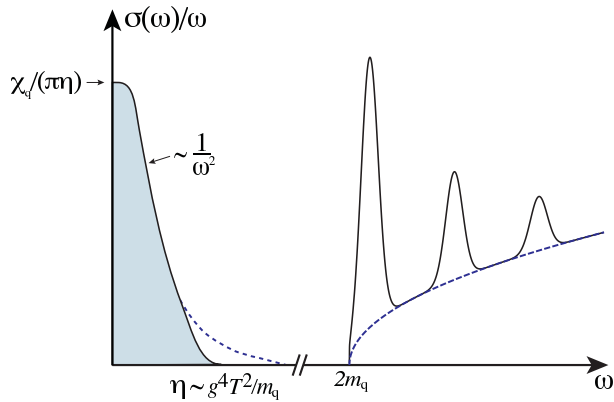}}
\caption{The schematic structure of the vector spectral function for heavy quarks. The arrow indicates
the value of the spectral function at zero frequency.} 
\label{fig:spf_demo}
\end{figure}

Since the analysis of the quarkonium spectral functions using MEM turns out to be
quite complicated,  it is important to
understand the temperature-dependence of the quarkonium correlation functions,
as well as try to identify possible sources of the temperature-dependence that are related
to melting of the bound states. From Eq. (\ref{eq.kernel}) it is clear that the temperature-dependence of the meson correlation functions comes from two sources: the trivial
temperature-dependence of the integration kernel $K(\tau,\omega)$ and the temperature-dependence of the spectral function $\sigma(\omega,T)$. To get rid of the first trivial
temperature-dependence, one can consider the reconstructed correlation function 
\beq
G_{rec}(\tau,T) = \int_0^{\infty} d \omega \sigma(\omega,T=0) K(\omega, \tau).
\eeq
If the spectral function does not change across the deconfinement transition 
$G(\tau,T)/G_{rec}(\tau,T)$ should be unity. Deviations of this ratio from unity indicate
a temperature-dependence of the spectral functions. The ratio $G(\tau,T)/G_{rec}(\tau,T)$
was first studied in Ref.~\refcite{Datta:2003ww} and subsequently in Refs.~\refcite{Datta:2004im,Datta:2004js,Datta:2006ua}, and \refcite{Jakovac:2006sf}. 
It was found that in the pseudoscalar
channel this ratio stays close to one, showing only small temperature-dependence.
This result seemed to support the early conclusion based on the spectral functions extracted from MEM, that
ground state charmonium survives in the deconfined phase up to temperatures $1.6\,T_c~$.
In the scalar and axial-vector channels large temperature dependence in $G(\tau,T)/G_{rec}(\tau,T)$
was observed. This was interpreted as melting of 1P charmonium states and fit well into
the expected picture of sequential melting. However, as already mentioned above,  
this conclusion was premature. In the deconfined phase there is an additional contribution to the spectral functions
at very low frequency. This can be seen by calculating the spectral function in the free theory, where a contribution proportional to
$\omega \delta(\omega)$ appears in all but the pseudoscalar channel \cite{Aarts:2005hg}. This is addition to the
continuum contributions which start at $\omega=2 m_q$, where $m_q$ is the heavy quark mass. The delta function gets smeared once 
the interactions of the heavy quarks with the medium are taken into account. In particular,
in the vector channel, where this low $\omega$ structure is related to the heavy
quark diffusion we have \cite{Petreczky:2005nh}
\beq
\omega \delta(\omega) \rightarrow \frac{1}{\pi} \frac{\eta \omega}{\omega^2+\eta^2}, ~\eta=T/(D m_q),
\eeq
$D$ being the heavy quark diffusion constant.
The schematic structure of the spectral functions is shown in Fig. \ref{fig:spf_demo}.
The area under the transport peak at low $\omega$ is given by the quark number susceptibility $\chi_q$.
For large quark mass the transport peak is narrow and is well-separated 
from the high energy part of the spectral function, which corresponds to bound states and/or unbound heavy quark anti-quark
pairs (continuum). One expects a similar structure in the spectral function in other channels as well.
Thus we can write
\ber
\sigma(\omega,T)&=&\sigma_{low}(\omega,T)+\sigma_{high}(\omega,T),\\
G(\tau,T)&=&G_{low}(\tau,T)+G_{high}(\tau,T).
\eer
Since the peak at low $\omega$ is narrow, we expect that the derivative of
$G_{low}(\tau,T)$ with respect of $\tau$ is small, $G_{low}'(\tau,T) \simeq 0$.
Lattice calculations in quenched QCD show that the ratios $G'(\tau)/G_{rec}'(\tau)$ 
have no strong temperature-dependence 
and are close to one for all temperatures, including the highest temperature of $3\,T_c$ 
\cite{Petreczky:2008px}. This means that the large temperature-dependence
seen in $G/G_{rec}$ for the scalar and axial-vector channels 
comes mostly from the temperature-dependence of $G_{low}(\tau,T)$ and is
not related to melting of P-wave  charmonium states. 
At the same time $G_{high}$, which is sensitive to the bound state properties, does not show
significant temperature-dependence for all of the channels.
A question thus arises, as to how the observed temperature-(in)dependence of $G_{high}(\tau,T)$ is related to the expected in-medium
modification of the heavy quark potential and the melting of quarkonium states at sufficiently high temperatures. This question is addressed below in context of potential models in Sections \ref{label:potmodel1} and \ref{label:potmodel2}.

\subsubsection{Spatial Meson Correlators}

In lattice QCD one also calculates a meson correlation function in one of the spatial
directions, say $z$
\begin{equation}
G(z,T)=\int dx dy \int_0^{1/T} d \tau \langle J(x,y,z,\tau) J(0,0,0,0) \rangle .
\end{equation}
The spatial correlation function is related to the meson spectral function
at non-zero spatial momentum
\begin{equation}
G(z,T)=\int_{-\infty}^{\infty} d p_z e^{i p_z z} \int_{0}^{\infty} d \omega 
\frac{\sigma(\omega,p_z,T)}{\omega}.
\end{equation}
Thus the temperature-dependence of the spatial correlation function 
also provides
information about the temperature-dependence of the spectral function.
Unlike the temporal correlators, spatial correlators can be calculated
for arbitrarily large separation and, therefore, can be more sensitive to in-medium
modification of quarkonium spectral functions.
Medium effects are expected to be the largest at distances which are larger than $1/T$.
At these distances $G(z,T)$ decays exponentially. This exponential decay is governed
by the so-called screening mass, $M_{scr}~$. If there is a lowest lying meson state of mass 
$M$, i.e. the spectral function can be well-approximated by Eq. (\ref{eq.stable}),
then the long-distance behavior of the spatial meson correlation 
function is determined by the meson mass, i.e. $M_{scr}=M~$. At very high temperatures, the quark and anti-quark 
are not bound and the meson screening mass is given by $2 \sqrt{(\pi T)^2+m_q^2}~$, 
where $m_q$ is the quark mass and $\pi T$ is the lowest Matsubara frequency. 
Therefore, a detailed study of spatial meson correlators
and screening masses can provide some information about the melting of meson states at high
temperatures. 
Spatial meson correlators and screening masses have been calculated in the light hadron sector \cite{Cheng:2010fe}
and for charmonium \cite{Karsch:2012na}.
The changes in the correlators and screening masses around the transition
temperature are much smaller for quarkonia than for the light mesons. This fits into the picture that ground state
of charmonium does not melt at the crossover temperature. However, for $T> 200~$MeV we see large change in spatial
charmonium correlators at distances $z T>1/2~$. Furthermore, for $T>300~$MeV
the behavior of the screening masses and their dependence on
the spatial boundary conditions is compatible with the picture of unbound quarks \cite{Karsch:2012na}.
This corroborates
the discussion in the previous subsection that there is no evidence for survival of charmonium states
at temperatures of $1.6\,T_c~$. 

\subsection{Effective Field Theory Approach}
\label{sec_pnrqcd}

There are different scales in the heavy quark bound state problem related to the heavy quark mass $m_q$,
the inverse size $\sim m_q v \sim 1/r~$, and the binding energy $~m_q v^2 \sim \alpha_s/r~$. Here $v$ is the 
typical heavy quark velocity
in the bound state and is considered to be a small parameter.
Therefore, it is possible to derive a sequence of effective field theories using this 
separation of scales (see Refs.~\refcite{Brambilla:2004jw} and \refcite{Brambilla:2010cs} for  recent reviews). 
Integrating out modes at the highest energy scale $\sim m_q$ leads to
an effective field theory called non-relativistic QCD (NRQCD), where the pair creation of heavy quarks is
suppressed by powers of the inverse mass and the heavy quarks are described by non-relativistic Pauli 
spinors \cite{Caswell:1985ui}.
At the next step, when the large scale related to the inverse size is integrated out, the potential NRQCD
or pNRQCD appears. In this effective theory the dynamical fields include the singlet 
$\rm S(r,R)$ and octet $\rm O(r,R)$ fields 
corresponding to the heavy quark anti-quark pair in singlet and octet states respectively, 
as well as light quarks and gluon fields
at the lowest scale $\sim m_q v^2$. The Lagrangian of this effective field theory has the form
\begin{eqnarray}
{\cal L} =
&&
- \frac{1}{4} F^a_{\mu \nu} F^{a\,\mu \nu}
+ \sum_{i=1}^{n_f}\bar{q}_i\,iD\!\!\!\!/\,q_i
+ \int d^3r \; {\rm Tr} \,
\Biggl\{ {\rm S}^\dagger \left[ i\partial_0 + \frac{\nabla_r^2}{m_q}-V_s(r) \right] {\rm S}\nonumber\\
&&
+ {\rm O}^\dagger \left[ iD_0 + \frac{\nabla_r^2}{m_q}- V_o(r) \right] {\rm O} \Biggr\}
+ V_A\, {\rm Tr} \left\{  {\rm O}^\dagger {\vec r} \cdot g{\vec E} \,{\rm S}
+ {\rm S}^\dagger {\vec r} \cdot g{\vec E} \,{\rm O} \right\}
\nonumber\\
&&
+ \frac{V_B}{2} {\rm Tr} \left\{  {\rm O}^\dagger {\vec r} \cdot g{\vec E} \, {\rm O}
+ {\rm O}^\dagger {\rm O} {\vec r} \cdot g{\vec E}  \right\}  + \dots\;.
\label{pNRQCD}
\end{eqnarray}
Here the dots correspond to terms which are higher order in the multipole expansion \cite{Brambilla:2004jw}.
The relative distance $r$ between the heavy quark and anti-quark plays a role of a label, the light
quark and gluon fields depend only on the center-of-mass coordinate $R$. The singlet $V_s(r)$ and octet $V_o(r)$ 
heavy quark potentials
appear as matching coefficients in the Lagrangian of the effective field theory
and, therefore, can be rigorously defined in QCD at any order of the perturbative expansion.
At leading order 
\beq
V_s(r)=-\frac{4}{3} \frac{\alpha_s}{r},~V_o(r)=\frac{1}{6}\frac{\alpha_s}{r}
\eeq
and $V_A=V_B=1$.
One can generalize this approach to finite temperature. However, the presence of additional 
scales makes the analysis more complicated \cite{Brambilla:2008cx}. The effective Lagrangian will have the same form as above,
but the matching coefficients may be temperature-dependent. In the weak coupling regime there are three different thermal scales:
$T$, $g T$ and $g^2 T$. The calculations of the matching coefficients depend on the relation of these thermal scales to
the heavy quark bound-state scales \cite{Brambilla:2008cx}. To simplify the analysis the static approximation has been used, in which
case the scale $m_q v$ is replaced by the inverse distance $1/r$ between the static quark and anti-quark. The binding energy
in the static limit becomes $V_o-V_s \simeq N \alpha_s/(2 r)$. When the binding energy is larger than the temperature the 
derivation of pNRQCD proceeds
in the same way as at zero temperature and there are no medium modifications of 
the heavy quark potential \cite{Brambilla:2008cx}. 
But bound state
properties will be affected by the medium through interactions with ultra-soft gluons.  In particular, 
the binding energy will be reduced
and a finite thermal width will appear due to medium-induced singlet-octet transitions arising from the dipole interactions in
the pNRQCD Lagrangian \cite{Brambilla:2008cx} (c.f. Eq. (\ref{pNRQCD})).
When the binding energy is smaller than one of the thermal scales, the singlet
and octet potential will be temperature-dependent and will acquire an imaginary part \cite{Brambilla:2008cx}. 
The imaginary part of the potential arises because of the singlet-octet transitions induced by the dipole vertex as well as
due to the Landau damping in the plasma, i.e. scattering of the gluons with space-like momentum off the thermal excitations in
the plasma. 
In general, the thermal corrections
to the potential go like $(r T)^n$ and $(m_D r)^n$ \cite{Brambilla:2008cx}, where $m_D$ denotes the Debye mass. 
Only for distances $r>1/m_D$ there is an exponential screening. 
In this region the singlet potential has a simple form
\begin{equation}
V_s(r)=
 -\frac{4}{3} \,\frac{\als}{r}\,e^{-m_Dr}
+ i \frac{4}{3}\,\als\, T\,\frac{2}{rm_D}\int_0^\infty dx \,\frac{\sin(m_Dr\,x)}{(x^2+1)^2}-\frac{4}{3}\, \als \left( m_D + i T \right),\nn\\
\label{Vs}
\end{equation}
The real part of the singlet potential coincides with the leading order 
result for the  free energy \cite{Petreczky:2005bd}.
The imaginary part of the singlet potential in this limit was first calculated in Ref.~\refcite{Laine:2006ns}.
For small distances, the imaginary part vanishes, while at large distances it is twice the damping rate of a
heavy quark \cite{Pisarski:1993rf}. This fact was first noted in Ref.~\refcite{Beraudo:2007ky} for thermal QED.
The effective field theory at finite temperature
has been derived in the weak-coupling regime assuming the separation of the 
different thermal scales and also $\Lambda_{QCD}$.
In practice, the separation of these scales is not evident and 
one needs lattice techniques to test this approach.

\subsubsection{Weak Coupling Results on Quarkonium Properties}

The binding energies and the thermal width of quarkonium have been calculated using pNRQCD in
the weak coupling regime, assuming the following hierarchy of scales: $m_q \alpha_s \gg T \gg m_q \alpha_s^2 \gg m_D$.
\cite{Brambilla:2010vq}. This regime may be of relevance for the $\Upsilon$, the ground state bottomonium in the QGP.
It was found that the thermal shift in the binding energy is proportional to $T^2/(\alpha_s m_q)$, while the thermal
width is proportional to $\alpha_s T$. In Ref.~\refcite{Brambilla:2011sg} 
it has been shown that the thermal width obtained in effective field theory framework naturally incorporates previous estimates of the width 
that rely on kinetic theory and the gluo-dissociation
cross-section \cite{Bhanot:1979vb}. The structure of the spin-dependent correction to the finite
temperature potential has been studied in the weak coupling regime \cite{Brambilla:2011mk}, as well as the 
thermal correction to the potential for bound states that move with respect to the rest frame of the plasma \cite{Escobedo:2011ie}.
In particular, it was found that above certain temperature the Landau-damping is not effective as a mechanism
for bound state dissociation \cite{Escobedo:2011ie}.

\subsubsection{Beyond Weak Coupling: Potential Models}
\label{label:potmodel1}

At increasingly high temperates quarkonium binding energies are significantly reduced from their vacuum values and eventually vanish. This is called the point of zero binding. In this situation the binding energy is the smallest scale in the problem and all the medium effects can be incorporated in the potential. If we start from pNRQCD and neglect the octet-singlet interactions the 
dynamics of the singlet quark anti-quark field is given by the free field
equation, which is the form of the Schr\"odinger equation (e.g. see discussions in
Ref.~\refcite{Brambilla:2004jw}). Therefore, the calculation of the correlation function
of singlet fields in real time, or equivalently the calculation of quarkonium spectral function
in the non-relativistic limit, is reduced to calculating the Greens-function of the Schr\"odinger equation \cite{Burnier:2007qm}
\begin{eqnarray}
&
\displaystyle
\left [ -\frac{1}{m_q} \vec{\nabla}^2+V(r,T)-E \right ] G^{nr}(\vec{r},\vec{r'},E,T) = \delta^3(r-r')
\nonumber\\[2mm]
&
\displaystyle
\sigma(\omega,T)=\frac{6}{\pi} {\rm Im} G^{nr}(\vec{r},\vec{r'},E)|_{\vec{r}=\vec{
r'}=0}\,~~~~ E=\omega-2 m.
\label{schroedinger}
\end{eqnarray}
The potential entering the above equation is complex \cite{Brambilla:2008cx,Laine:2006ns}, and, as discussed above, its form is quite complicated even in the weakly coupled regime, since the form of the
thermal corrections depends on the relation of the scales $1/r$, $T$, $m_D \sim g T$.
Furthermore, in the relevant temperature regime the separation of the above scales does not
hold and the potential is effected by the non-perturbative scales, such as $g^2 T$ and $\Lambda_{QCD}~$. Therefore, in this case we 
have to rely on the lattice calculations to constrain the form of the potential.

\subsubsection{Potential Models and Euclidean Correlation Functions}
\label{label:potmodel2}

\begin{figure}
\includegraphics[width=6.3cm]{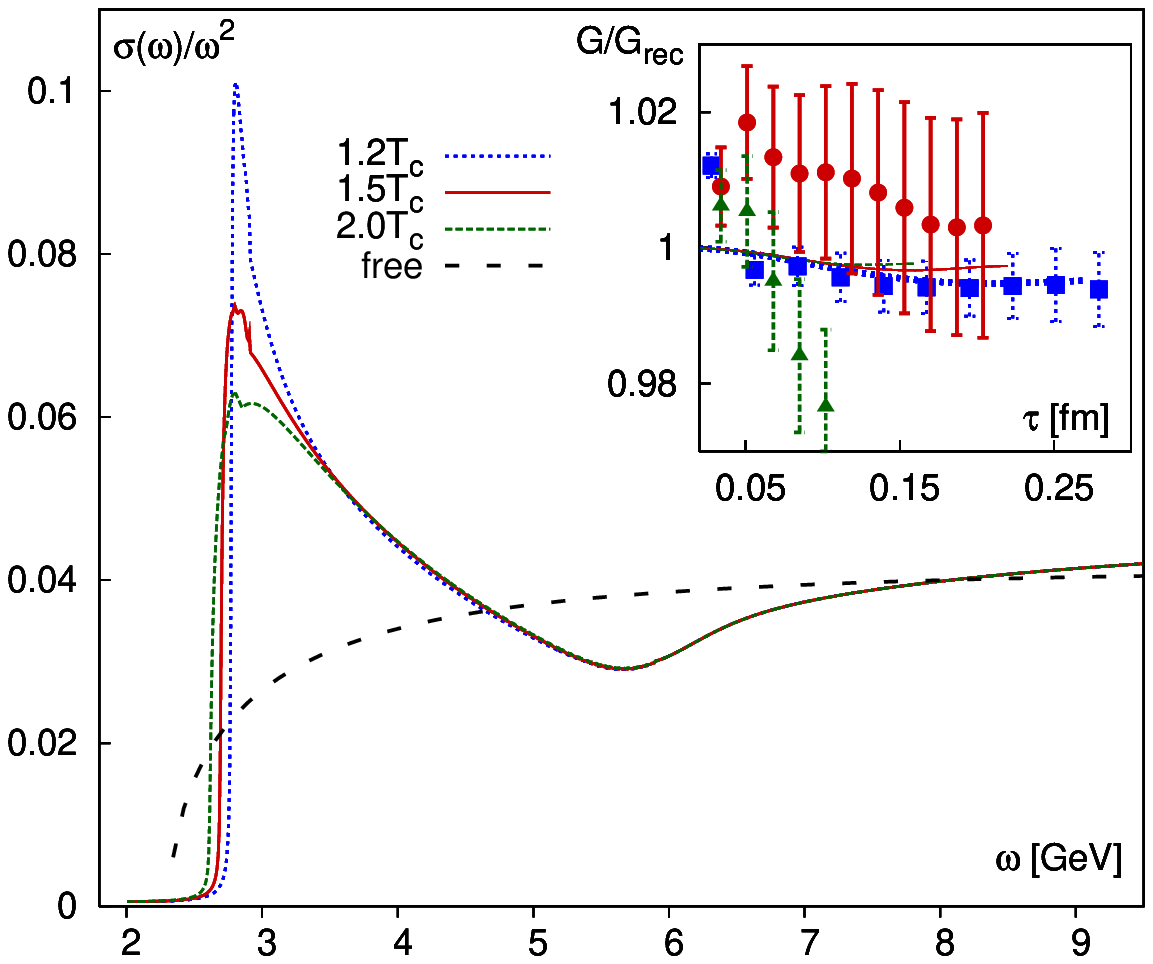}
\includegraphics[width=6.3cm]{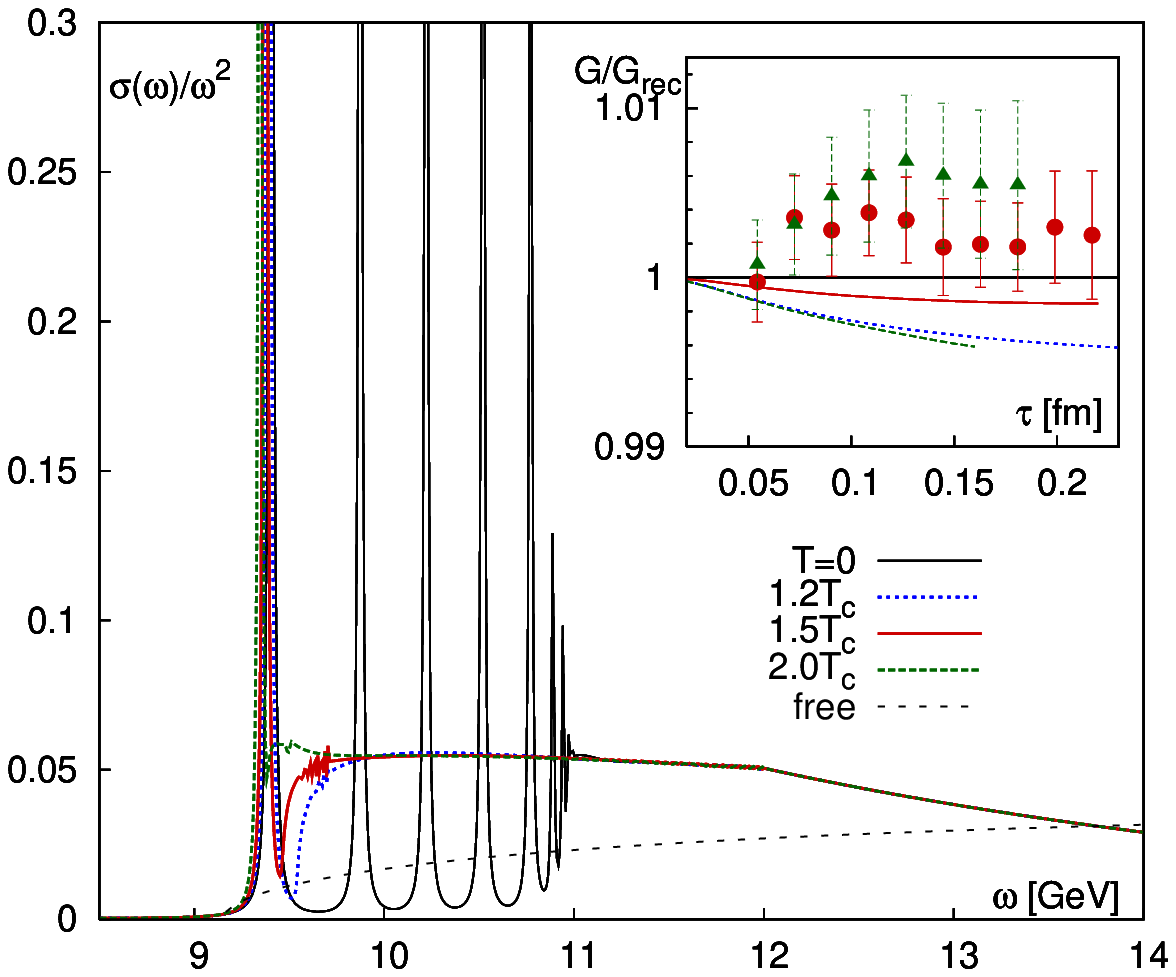}
\caption{The S-wave charmonium (left) and bottomonium (right) spectral functions calculated for
quenched QCD using a lattice QCD inspired potential \cite{Mocsy:2007yj}. The insets show the ratio
$G/G_{rec}$ obtained from the calculated spectral functions and compared to lattice QCD data.}
\label{fig:spf1}
\end{figure}

A simple potential model based on lattice QCD data on the singlet free energy has been
proposed to study quarkonium spectral functions and explain the observed temperature-independence of the quarkonium correlators \cite{Mocsy:2007yj,Mocsy:2007jz}. The imaginary part of the potential
has been neglected in this work, while the real part of the potential was based on lattice QCD data
on the singlet free energy. In this model all quarkonium states except the ground state bottomonium melt in the QGP.
However, the interaction between the heavy quark and anti-quark leads to large enhancement of the spectral function in
the threshold region. This is shown in Fig. \ref{fig:spf1} for the case of S-wave charmonium spectral function in
quenched QCD. One can see, that there is no bound
state already at $1.2\,T_c$, while the correlation functions do not show large temperature-dependence. The ratio 
$G/G_{rec}$ is close to unity, in agreement with the lattice findings. Similar results have been found also for the P-wave quarkonium \cite{Mocsy:2007yj,Mocsy:2007jz}. This result is understood in that the large threshold enhancement compensates for the absence of
bound states and therefore the correlators do not change much. Even though the difference between the vacuum and in-medium
spectral function increases with increasing temperatures, the Euclidean time extent $1/(2T)$ becomes smaller. Therefore,
it is difficult to see this change in Euclidean time correlators. Similar results have been reported also in Ref.~\refcite{Alberico:2007rg}.
Furthermore, the above findings do not depend on the precise choice of the potential, provided it is compatible with lattice QCD
findings at short distances \cite{Mocsy:2007yj,Mocsy:2007jz} (see also discussions in Ref.~\refcite{Mocsy:2007py}). This analysis resolved the apparent puzzle between the strong modification of the potential and very small temperature dependence of the Euclidean quarkonium correlators.

\subsubsection{Static Energy in Lattice QCD}

\begin{figure}
\includegraphics[width=6.3cm]{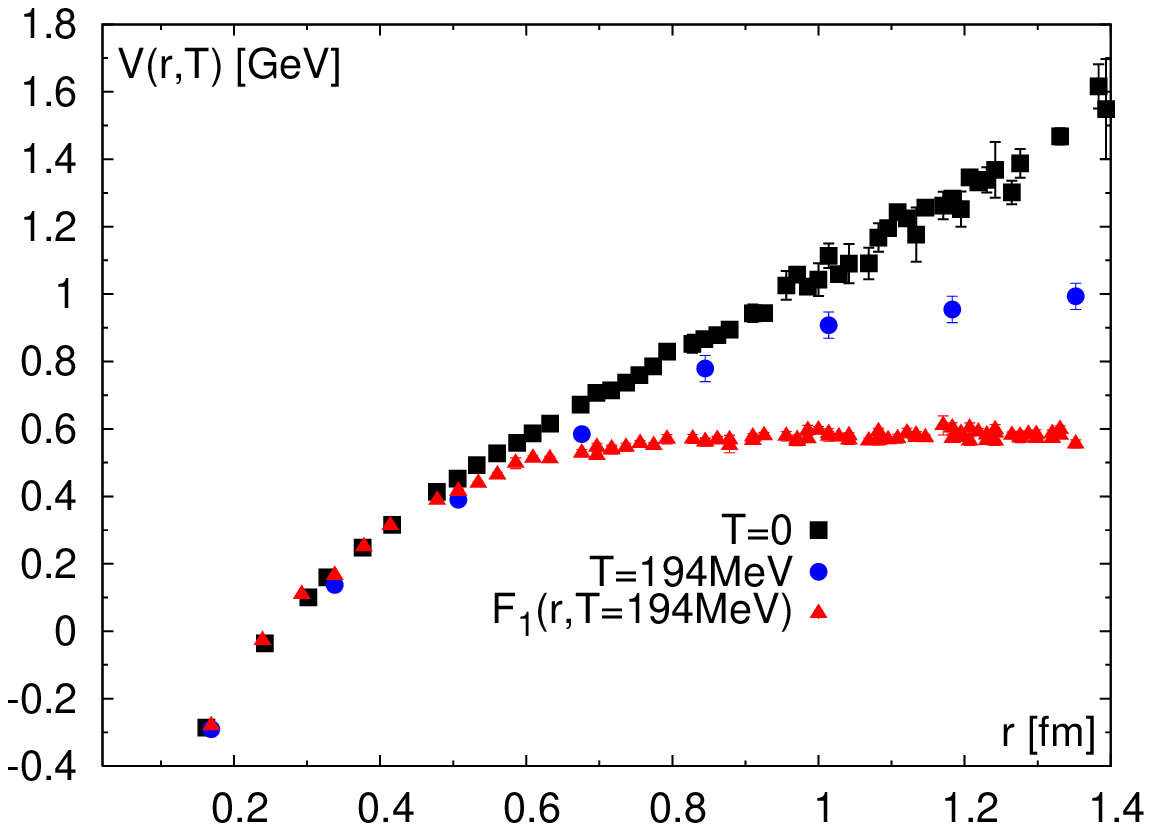}
\includegraphics[width=6.3cm]{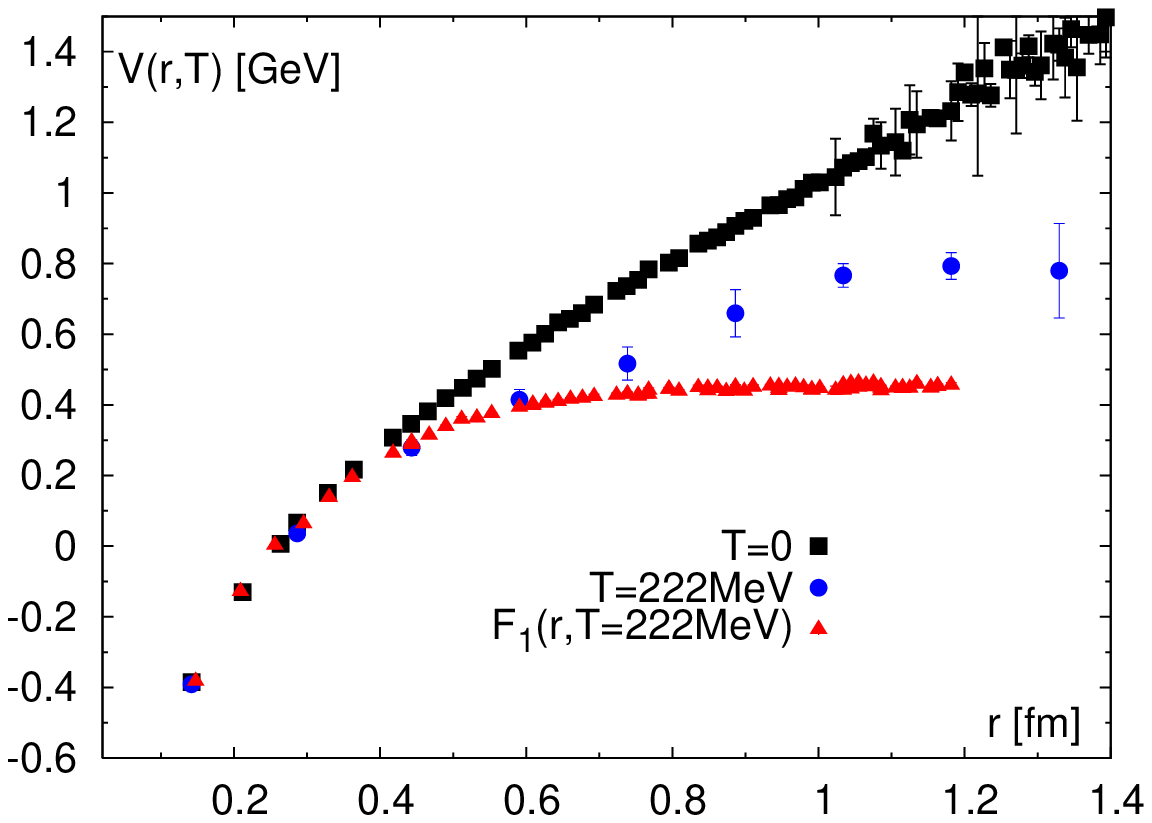}
\vspace*{-0.4cm}
\caption{
The static quark anti-quark energy extracted for $T=178~$MeV (left) and $T=220~$MeV (right) compared to the zero temperature result and to the singlet free energy.
}
\label{fig:wloop}
\end{figure}

As already mentioned, in the limit of small binding energy all the thermal scales, as well as $\Lambda_{QCD}~$, 
can be integrated out, and
the potential can be approximated by the static energy. The static energy has been calculated in
perturbation theory \cite{Brambilla:2008cx} and is complex-valued. Since the validity of perturbation 
theory is not evident for the relevant temperature range, it is interesting to calculate
the static energy on the lattice.
In the previous section we considered the correlation function of a static $Q\bar Q$ pair evaluated
at Euclidean time $t=1/T$. These correlators are related to the free energy of a static $Q\bar Q$
pair. One can consider Wilson loops for Euclidean times $t < 1/T$, which have no obvious relation
to the free energy of a static $Q\bar Q$ pair. Wilson loops at non-zero temperature have been
first studied in Refs.~\refcite{Rothkopf:2009pk} and \refcite{Rothkopf:2011db} in connection with the heavy quark potential at non-zero temperature.  
A spectral decomposition for the Wilson loops has been conjectured: 
\beq
W(r,\tau)=\int_0^{\infty} \sigma(\omega,r,T) e^{-\omega \tau}.
\eeq
At zero temperature, the spectral function is proportional to a sum of delta functions $\sigma(r,\omega)=\sum_n c_n \delta(E_n(r)-\omega)~$.
At high temperatures, the spectral function is proportional to a sum of smeared delta functions. The position and width of the 
lowest peak in the spectral function are related to real and imaginary part of the potential, respectively \cite{Rothkopf:2009pk,Rothkopf:2011db}. 
Motivated by this, Wilson loops have been calculated on finite temperature lattices in 2+1 flavor QCD using the HISQ action with physical
strange quark mass and light quark masses that correspond to a pion mass of $160$ MeV in the
continuum limit \cite{Bazavov:2012bq,Bazavov:2012fk}. From the analysis
of the Wilson loops, the real part of the static energy of $Q\bar Q$ pair has been extracted. 
Unfortunately, the Wilson loops are not very sensitive to the imaginary part of the static energy.
The corresponding results
at two temperatures are shown in Fig. \ref{fig:wloop}. The real part of the static energy is larger than the singlet
free energy. As one can see from the figure, screening effects are not present at $T=178~$MeV but can be clearly
seen for $220~$MeV. This finding is in accord with the expectations from the temperature-dependence
of $F_{\infty}(T)$ which seems to be described well by hadronic degrees of freedom for $T< 200~$MeV (see the previous section).

\begin{figure}
\includegraphics[width=6.3cm]{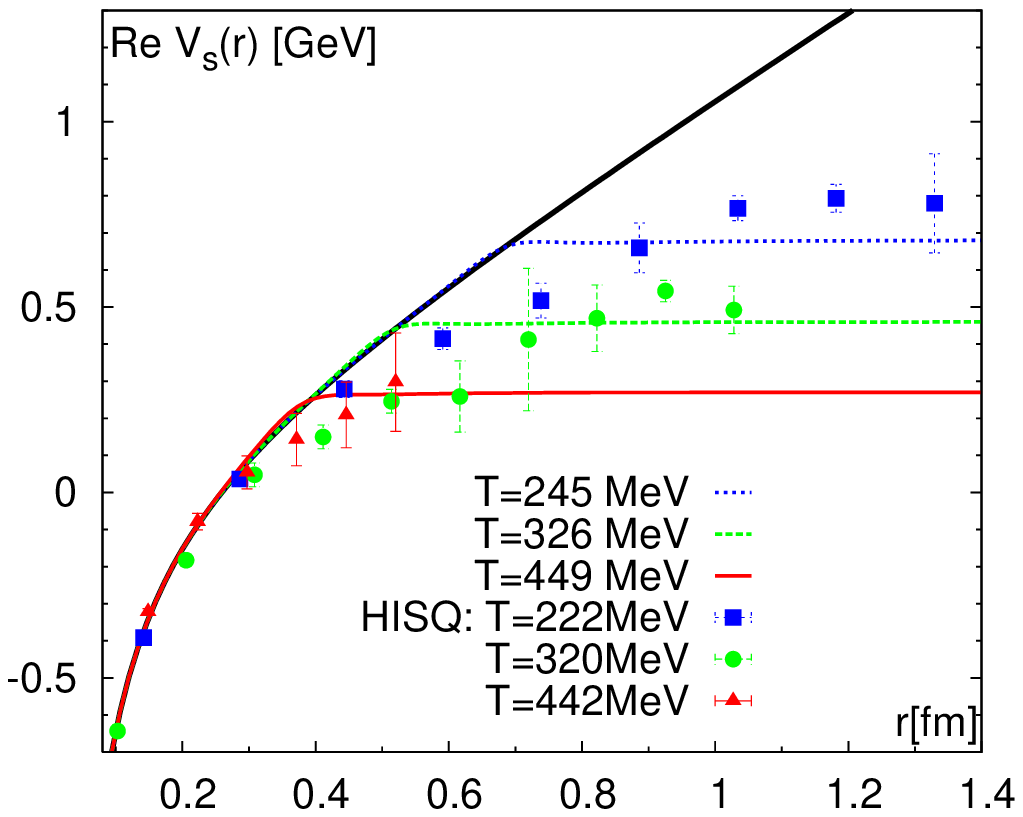}
\includegraphics[width=6.3cm]{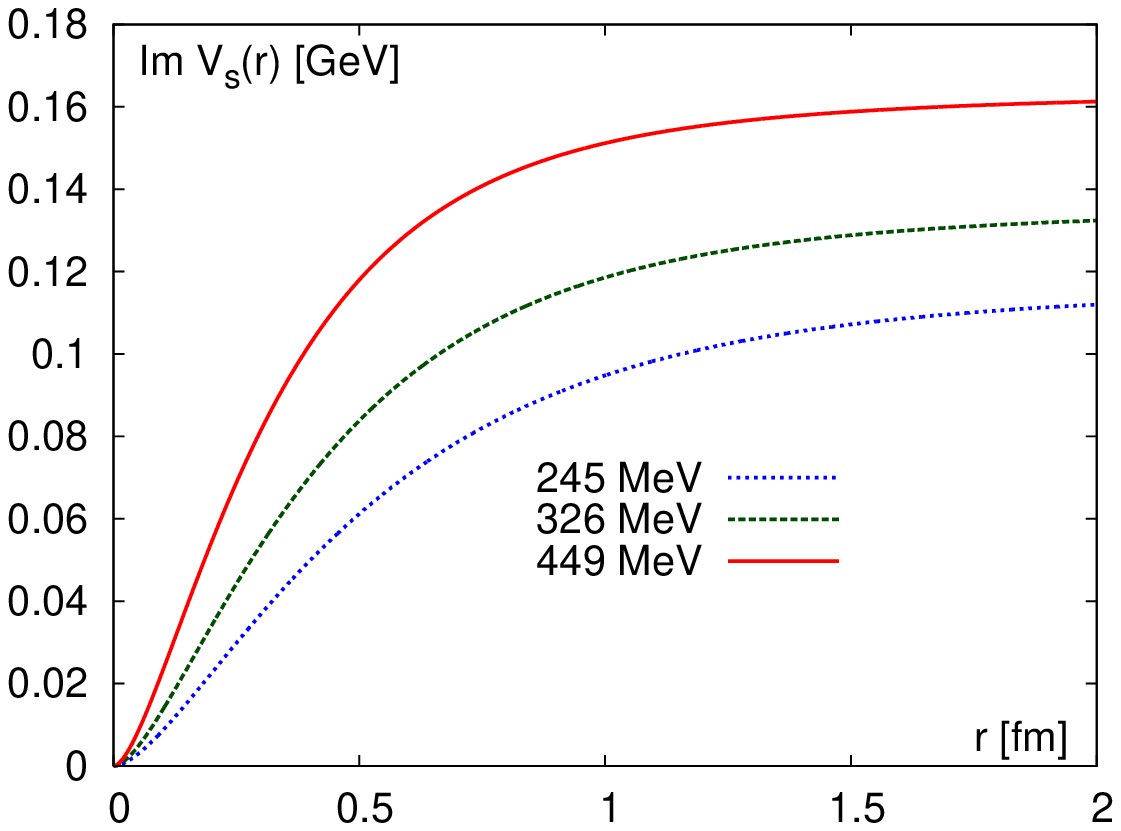}
\caption{The real (left) and imaginary (right) part of the potential used in the calculation of
quarkonium spectral functions \cite{Petreczky:2010tk}. Also shown are lattice results
on the static energy extracted from Wilson loops \cite{Bazavov:2012bq}.}
\label{fig:pot}
\end{figure}

\subsubsection{Spectral Functions in a Realistic Potential Model}

\begin{figure}
\includegraphics[width=6.3cm]{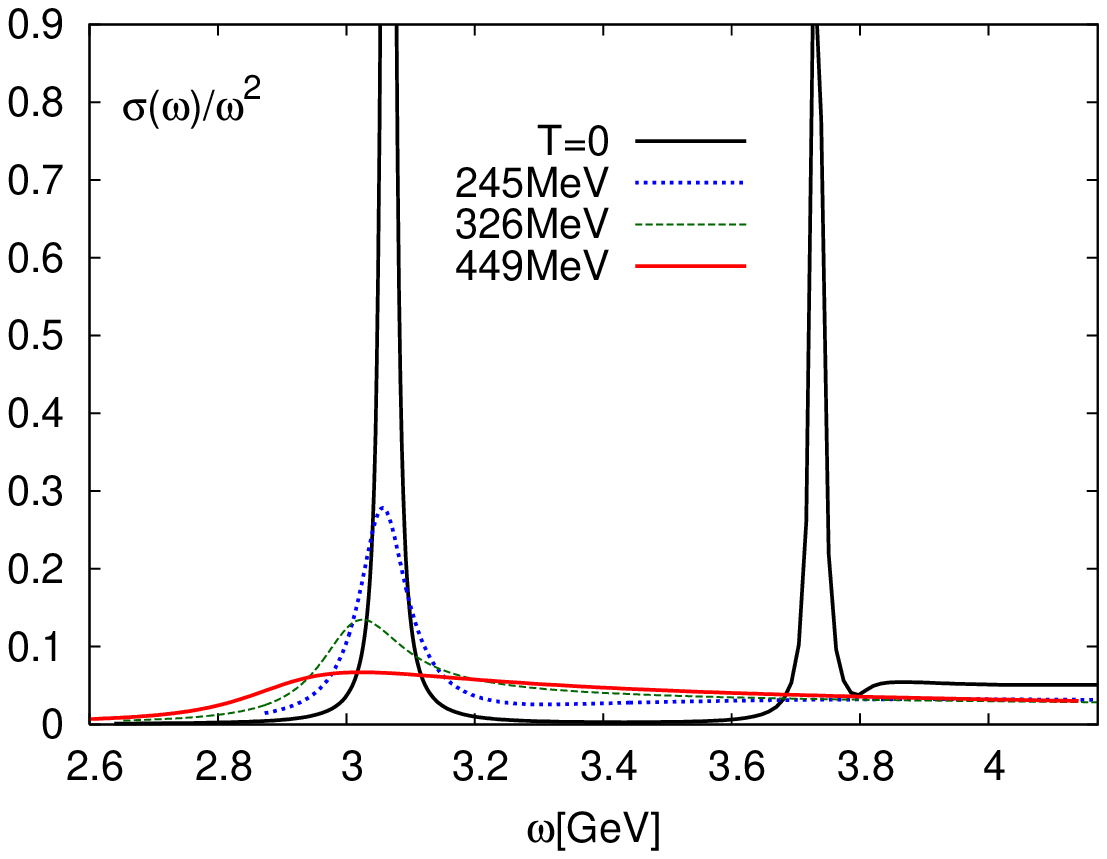}
\includegraphics[width=6.3cm]{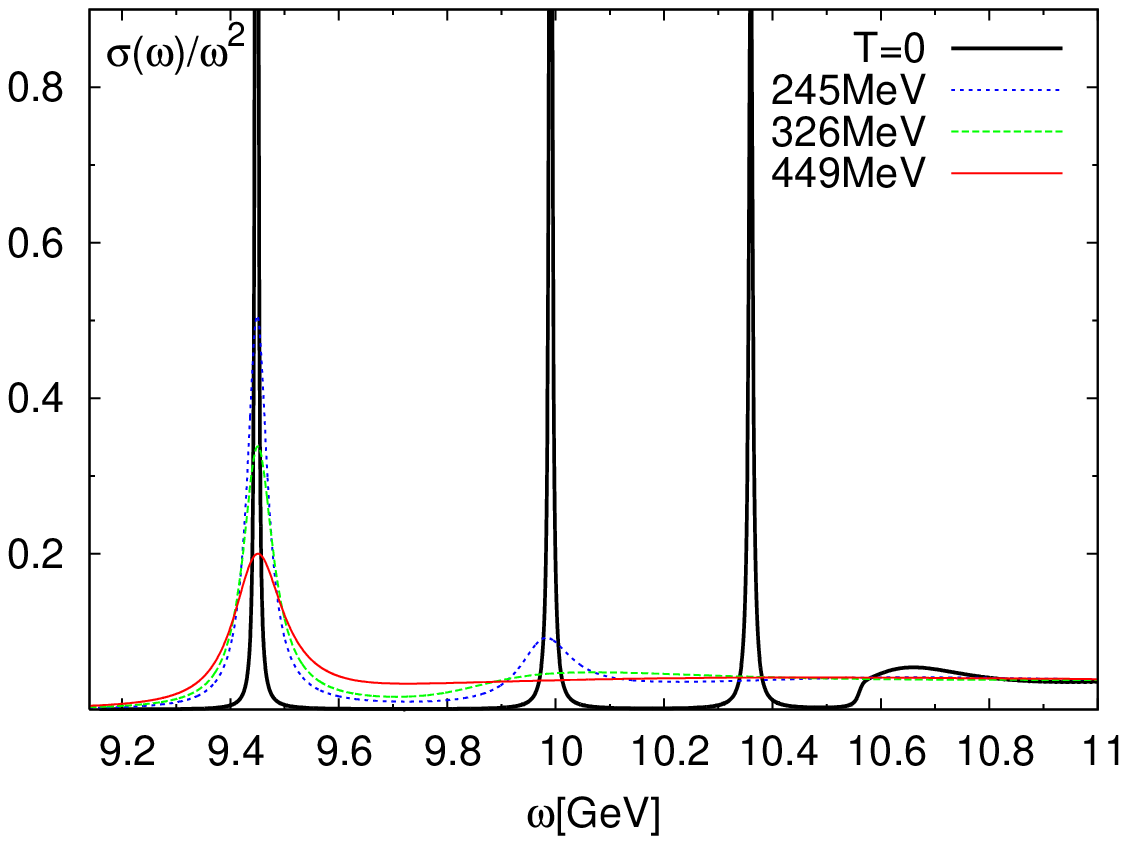}
\caption{Charmonium (left) and bottomonium (right) spectral functions calculated in a potential model
with complex potential \cite{Petreczky:2010tk}.}
\label{fig:spf}
\end{figure}

As discussed above, the in-medium corrections to the potential have both real and imaginary parts. For very small couplings the imaginary part of the potential is actually the dominant source of bound state dissolution. Therefore, any realistic calculation of quarkonium spectral functions should include both real and imaginary parts for the temperature-dependent potential. The hierarchy of scales assumed
in weak coupling calculations may not be satisfied and non-perturbative effects may play an important
role (see previous section). Thus, one has to rely on the lattice results. But, as mentioned before, the imaginary part of the potential cannot be extracted from
present lattice calculations. Therefore, in Ref.~\refcite{Petreczky:2010tk} the imaginary part of the potential was evaluated according
to Eq. (\ref{Vs}) using the prescription for the temperature-dependent running coupling constant from Ref.~\refcite{Burnier:2007qm}.
For the real part of the potential lattice information on color screening was used. Namely, the potential
was exponentially screened for distances $r>r_{scr}=0.8/T$, and at smaller distances this was taken to be equal to the zero
temperature potential. This construction gives the maximal real part of the potential that still compatible with the data.
The real and imaginary part of the potential used in this analysis is shown in Fig. \ref{fig:pot}. We also show the static
energy calculated on the lattice for the matching values of the temperature. Interestingly, the real part of the potential
is close to the lattice data on the static energy. The calculated S-wave charmonium and bottomonium spectral functions are shown
in Fig. \ref{fig:spf}. As one can see from the figure, all of the quarkonium states, except the $1S$ bottomonium, are dissolved for
$T>250~$MeV. The $\Upsilon(1S)$ may survive up to temperatures of $450~$MeV and its melting will only occur at higher temperatures.
We would like to re-iterate, that the key ingredient in the observed melting pattern is the presence of the imaginary part of the potential; neglecting it
would lead to larger dissociation temperatures (see e.g. Ref.~\refcite{Zhao:2010nk}).
The above estimates of the melting temperatures are in agreement with earlier estimates presented in Ref.~\refcite{Mocsy:2007jz} (note, that $T_c=204~$MeV in those pure gauge calculations). Also, the above spectral functions agree qualitatively with the ones found in a $T$-matrix approach that also includes some of the effects of the imaginary potential \cite{Riek:2010py}.

\subsubsection{Lattice NRQCD at Non-zero Temperature}

For the most interesting cases the heavy quark mass is well-separated from other scales.
Therefore, NRQCD is always a good effective theory and can be combined with non-perturbative
approaches like lattice QCD.
A lattice version of NRQCD has been successfully used to study bottomonium spectral functions non-perturbatively
(see e.g. \cite{Gray:2005ur}). This approach can also be used to study bottomonium spectral
functions at non-zero temperatures \cite{Aarts:2010ek,Aarts:2011sm}. Studying bottomonium spectral functions using
relativistic heavy quarks turns out to be difficult because of the noise, which is not
a problem in the non-relativistic formulation. In the NRQCD approach heavy quarks are not
subject to anti-periodic conditions and, therefore, the meson correlator can be studied up to
twice larger distances $\tau<1/T$ and has a simpler spectral decomposition 
 \beq
G(\tau,T) = \int_{-\infty}^{\infty} d \omega \sigma(\omega,T) e^{-\omega \tau}.
\eeq
Due to the fact that gauge fields are still periodic, the spectral function can be non-zero
for negative energies \cite{Aarts:2011sm}.
In this approach there is no zero-mode contribution. This, and the fact that correlators can be studied at larger
distances, makes the correlators more sensitive to the medium-modification of the spectral functions. The temperature
dependence of P-wave bottomonium correlators appears to be strong at temperatures of about $280~$MeV, which is
indicative of melting of P states\cite{Aarts:2010ek}.  At a temperature of $400~$MeV, the behavior of the P-wave correlators is consistent
with propagation of unbound quarks\cite{Aarts:2010ek}.  In contrast, the S-wave bottomonium correlators do not show significant
temperature dependence which is consistent with survival of bottomonium states to temperatures $T \geq 400~$MeV.


\section{Quarkonium Phenomenology}

\subsection{The Anisotropic (Viscous) Quark Gluon Plasma}

State of the art dynamical models of QGP spatiotemporal evolution rely on some form of viscous hydrodynamical treatment.  Over a decade ago it was shown that ideal relativistic hydrodynamics was able to reproduce the soft collective flow of the matter and single particle spectra produced in relativistic heavy ion collisions.~\cite{Huovinen:2001cy,Hirano:2002ds,Tannenbaum:2006ch,Kolb:2003dz}  However, one expects on general grounds that there is a lower bound on the QGP shear viscosity.~\cite{Danielewicz:1984ww,Kovtun:2004de}. Based on this realization, there was subsequently a concerted effort to develop a systematic framework for describing the soft dynamics of the QGP using relativistic viscous (non-ideal) hydrodynamics.~\cite{Muronga:2001zk,Muronga:2003ta,Muronga:2004sf,Baier:2006um,Romatschke:2007mq,Schenke:2011tv,Niemi:2012ry,Bozek:2012qs,Denicol:2012cn}.

Traditional viscous hydrodynamics approaches assume that the system is close to thermal equilibrium and hence is also very close to being isotropic in momentum space.  However, one finds during the application of these methods that this assumption breaks down at the earliest times after the initial impact of the two nuclei, due to large momentum-space anisotropies in the $p_T$-$p_L$ plane, which can persist for the entire lifetime of the plasma \cite{Martinez:2009mf,Martinez:2010sd,Ryblewski:2010bs}. The level of the observed momentum-space anisotropy increases as the assumed shear viscosity of the QGP increases.  Similar conclusions have been obtained in the context of strongly coupled systems, where it has been shown that one achieves viscous hydrodynamical behavior at early times, however, the system still possesses large momentum-space anisotropies which persist throughout the evolution \cite{Chesler:2008hg,Chesler:2009cy,Heller:2011ju,Heller:2012je,Heller:2012km,Wu:2011yd,Chesler:2011ds}.  As a result, it is necessary to include the effect of momentum-space anisotropies on the screening of heavy quarkonium.

This is accomplished minimally by introducing a momentum-space anisotropy parameter, $\xi$, which is related to the ellipticity of the distribution function in the $p_T$-$p_L$ plane \cite{Romatschke:2003ms,Mrowczynski:2004kv,Romatschke:2004jh,Schenke:2006fz,Dumitru:2007hy}, into the underlying one-particle partonic distribution functions.  The state of the QGP is then described by a local average momentum scale, $\Lambda(\tau,{\bf x})$ (transverse temperature), and a local momentum anisotropy parameter $\xi(\tau,{\bf x})$.  If the system is close to being isotropic in momentum space, then $\xi$ can be mapped to the plasma shear, $\Pi$, and 2nd order viscous hydrodynamics is recovered; however, if the system is highly anisotropic it is possible take moments of the Boltzmann equation and derive a set of coupled partial differential equations.~\cite{Florkowski:2010cf,Martinez:2010sc,Ryblewski:2010bs,Martinez:2010sd,Ryblewski:2011aq,Martinez:2012tu,Ryblewski:2012rr,Florkowski:2012as}  The resulting dynamical equations allow one to vary the assumed value of the QGP shear viscosity in order to assess its impact on the dynamical evolution of the system.

Additionally, one must revisit the calculation of the heavy quark potential in the presence of momentum-space anisotropies.  This was first considered in Ref.~\cite{Dumitru:2007hy} in which the real part of the heavy quark potential was calculated.  Shortly afterwards, the imaginary part of the heavy quark potential in the the presence of momentum-space anisotropies was calculated \cite{Burnier:2009yu,Dumitru:2009fy,Philipsen:2009wg} and phenomenological potential models incorporating this effect were constructed \cite{Dumitru:2009ni,Margotta:2011ta,Strickland:2011aa}.  The chief effect of momentum-space anisotropies is to reduce Debye screening in the plasma.\footnote{Another interesting, but subleading, effect is that momentum-space anisotropies cause a splitting between $\ell\neq0$ states.}  This in turn has the effect that heavy quarkonium states can survive to higher temperatures in the presence of momentum-space anisotropies. The effect of mild ($\xi=1$) momentum space anisotropies is indicated in the right column of Table \ref{tab:dissociation}. From this we can see that, for all states shown, the dissociation temperature increases as the level of momentum-space anisotropy increases. We consider a state dissociated when its width is larger than its binding energy, because a state then decays faster than it binds. The temperatures shown in \ref{tab:dissociation} are the temperatures where thermal medium effects of $\sim T$ are comparable with the binding energy. This condition at finite temperatures is more realistic than the zero-binding condition, often quoted in the literature, which is sensible only at zero temperature.  

\begin{figure}[t]
\includegraphics[width=0.48\textwidth]{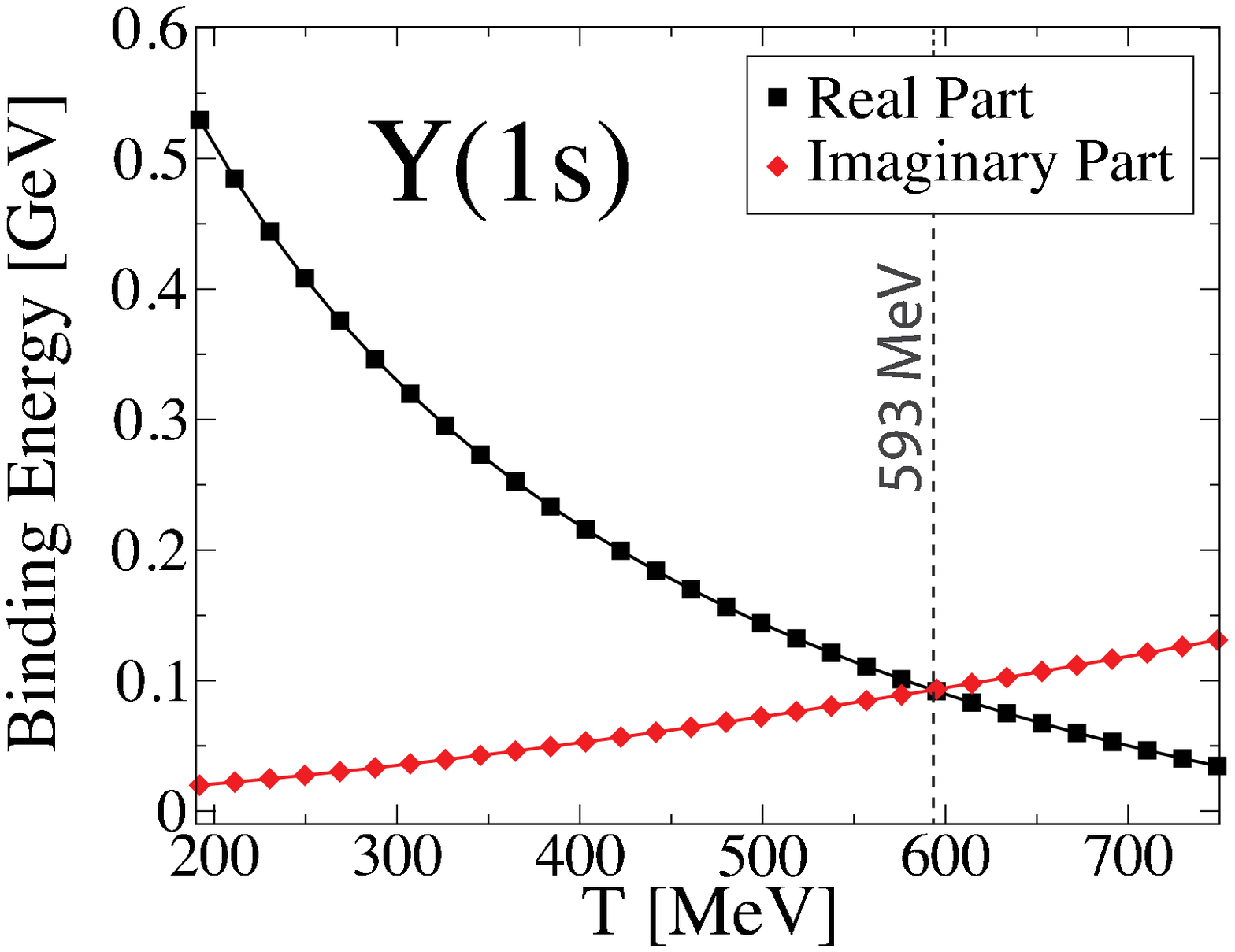}
\hspace{3mm}
\includegraphics[width=0.48\textwidth]{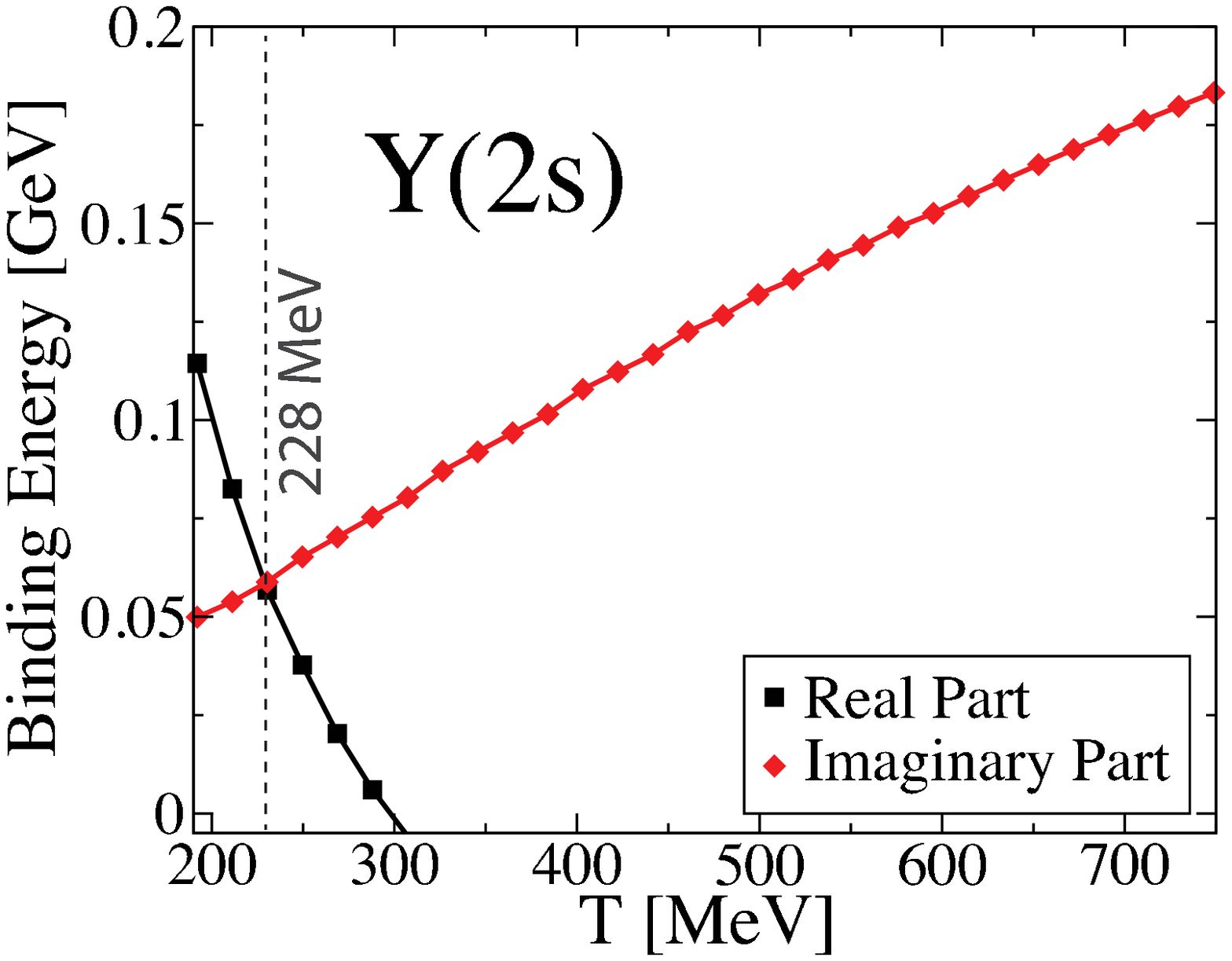}
\caption{Real and imaginary parts of the binding energy of the $\Upsilon(1s)$ (left) and $\Upsilon(2s)$ (right) states as a function
of the temperature for a momentum-space isotropic QGP.  Dashed vertical lines indicate the temperature at which the real and imaginary parts are equal.  Numerical results taken from Ref.~\protect\refcite{Strickland:2011aa}.  
}
\label{fig:upsilon-bind}
\end{figure}

\subsection{A Phenomenological Potential for an Anisotropic Plasma}

In Ref.~\refcite{Strickland:2011aa} a complex-valued potential model was constructed in order to address the question of $\Upsilon$
suppression, in particular.  The imaginary contribution to the potential was taken to be given by the leading order perturbative result~\cite{Laine:2006ns,Burnier:2009yu,Dumitru:2009fy}.  The real part was taken to be given by either the free energy or internal energy, with the free energy, $F$, given by a Karsch-Mehr-Satz potential form \cite{Karsch:1987pv} and the internal energy given by $U = F - T S$.  In both
cases, lattice results were used to fix empirical constants such as the string tension.  For simplicity, herein we focus on the results obtained using the more binding internal energy model, since the free-energy-based model seems to give too much suppression at both RHIC and LHC energies.  

\begin{table}[t]
\tbl{Estimates of the isotropic and anisotropic dissociation scales for the $J/\psi$, $\chi_{c1}$, $\Upsilon(1s)$, $\Upsilon(2s)$, $\Upsilon(3s)$, $\chi_{b1}$, and $\chi_{b2}$.  Estimates are taken from Refs.~\protect\refcite{Margotta:2011ta,Strickland:2011aa}.
}
{
\begin{tabular*}{.8\linewidth}{@{\extracolsep{\fill}}|l|l|l|}
\hline
{\bf State} & {\bf Isotropic QGP ($\xi$=0)} ~~ & {\bf  Anisotropic QGP ($\xi$=1)} ~~~ \\ \hline 
$J/\psi$ &  307 MeV &  374 MeV\\\hline
$\chi_{c1}$ & $<$ 192 MeV &  210 MeV \\\hline
$\Upsilon(1s)$ &  593 MeV &  735 MeV\\\hline
$\Upsilon(2s)$ & 228 MeV &  290 MeV \\\hline
$\Upsilon(3s)$ & $<$ 192 MeV &  $<$ 192 MeV  \\\hline
$\chi_{b1}$ & 265 MeV &  351 MeV \\\hline
$\chi_{b2}$ & $<$ 192 MeV &  213 MeV\\\hline
\end{tabular*}
}
\label{tab:dissociation}
\end{table}

In Fig.~\ref{fig:upsilon-bind} we show results for the real and imaginary parts of the binding energy of the $\Upsilon(1s)$ and the $\Upsilon(2s)$~\cite{Strickland:2011aa,Margotta:2011ta}.  If the real part of the binding energy is positive, then the state is bound.  If the real part of the binding energy is negative, then the state is unbound.  The imaginary part of the binding energy gives information about the decay rate of the state in question.  To see the exact relationship we can compute the quantum mechanical occupation number as a function of proper time
\begin{eqnarray}
n_\upsilon(\tau) &=& \langle \phi_\upsilon^*(\tau,{\bf x}) \phi_\upsilon(\tau,{\bf x}) \rangle \nonumber \, , \\
&=&  \langle\phi_\upsilon^*(\tau_0,{\bf x})\phi_\upsilon(\tau_0,{\bf x}) \rangle e^{2 \Im[E] (\tau-\tau_0)} \, , \nonumber \\
&=&  n^0_\upsilon \, e^{2 \Im[E] (\tau-\tau_0)} \, ,
\end{eqnarray}
where in the last line we have identified $n^0_\upsilon  = \langle\phi_\upsilon^*(\tau_0,{\bf x})\phi_\upsilon(\tau_0,{\bf x}) \rangle$.  In order to connect this to the decay rate, $\Gamma$, we note that $\Gamma$ is defined empirically through $n_\upsilon(t) = n^0_\upsilon \, \exp(-\Gamma (\tau-\tau_0))$ so that we can identify $\Gamma = -2 \Im[E]$.  As a rough guide one can identify a dissociation temperature with the point at which the real and imaginary parts of the binding energy become comparable.  At this point the state decays within one real-time oscillation of the quantum state.

In practice, however, the precise value of the disassociation temperature is of little importance since even below this temperature the state under consideration is undergoing in-medium decays.  In practice, one must compute the dynamical decay rate based on the local temperature and momentum-space anisotropy of the plasma and solve for the fraction of states remaining at a particular point in the plasma as a function of proper-time and then integrate/average over space-time.  In order to carry this step out it is necessary to have a model of the spatio-temporal dynamics of the QGP.  We will return this issue after the next subsection.

\subsection{Sequential Suppression and Feed Down}

As mentioned above, as rough guide one can determine a rough dissociation temperature for each state by finding the temperature at which the real and imaginary parts of the binding energies become equal.  As can be seen from Fig.~\ref{fig:upsilon-bind} (and Table \ref{tab:dissociation}), the dissociation temperature determined in this way is higher for the $\Upsilon(1s)$ state than it is for the $\Upsilon(2s)$ state meaning that the $\Upsilon(1s)$ will survive to higher temperatures.  In Table \ref{tab:dissociation} we list estimates of disassociation temperatures for the $J/\psi$, $\chi_{c1}$, $\Upsilon(1s)$, $\Upsilon(2s)$, $\Upsilon(3s)$, $\chi_{b1}$, and $\chi_{b2}$ states obtained in Refs.~\refcite{Margotta:2011ta,Strickland:2011aa}.  In those references the authors determined the real and imaginary parts of the binding energies for each of the states listed.  This is necessary since if, for example, one is interested in the suppression of the $J/\psi$ state, it is necessary to calculate the direct suppression of all states that can potentially decay into the $J/\psi$.

\begin{figure}[t]
\centerline{\includegraphics[width=0.97\textwidth]{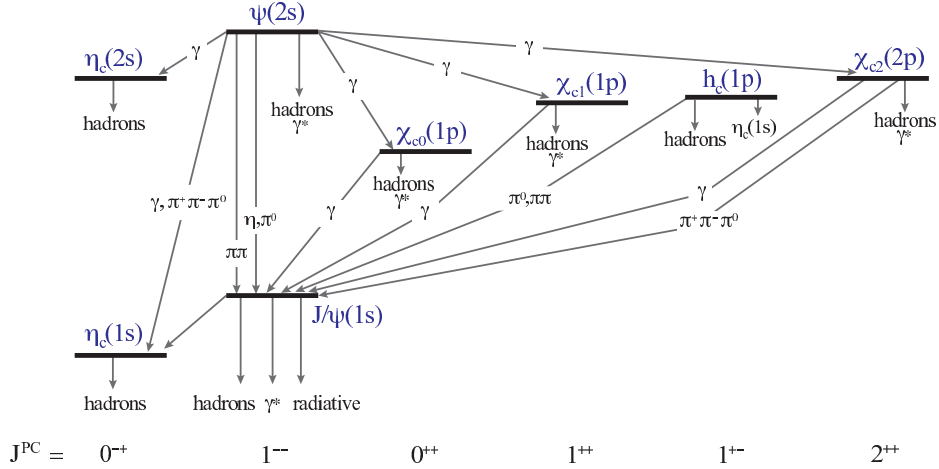}}
\caption{Spectroscopic diagram for charmonium states.  Decay channels shown were taken from Ref.~\protect\refcite{Beringer:1900zz}.
The bottom row shows the spin, parity, and charge conjugation quantum numbers  associated with the states above it.
}
\label{fig:charmonium-feeddown}
\end{figure}

The complications implied by feed down can be inferred from Fig.~\ref{fig:charmonium-feeddown} in which we have illustrated the spectrum of charmonium states along with the decay channels of each state~\cite{Beringer:1900zz}.  We do not show extremely rare decays or possible feed downs from bottom hadrons.  As can be seen from this figure, when computing $J/\psi$ suppression, for example, it is necessary to have information about the suppression of the $\Psi(2s)$, $\chi_{c\{1,2,3\}}(1p)$, and $h_c(1p)$ states.  A similar complication arises in the bottomonium sector.  In vacuum the percentage of produced, e.g., $J/\psi$ states which come from a particular decay channel is called the feed down fraction.  Many experiments have attempted measurements of the feed down fractions for the charmonium and bottomonium systems.  In Table \ref{tab:feeddown} we present feed down fractions for $J/\psi$  and $\Upsilon$ production.  The quoted numbers internal to the charmonium system are world averages presented in Ref.~\protect\refcite{Faccioli:2008ir}.  The $J/\psi$ bottom hadron ($b$-hadron) feed down fraction is estimated from Ref.~\protect\refcite{Acosta:2004yw}.  The direct $J/\psi$ production fraction is estimated by requiring that the sum of the listed feed down fractions be equal to one.  The bottomonium feed down fractions are quoted from Ref.~\protect\refcite{Affolder:1999wm}.  

As we can see from Table \ref{tab:feeddown}, in order to make quantitatively reliable calculations of quarkonium suppression, it is necessary to calculate the suppression of the first few excited states.  The direct suppression of each state (excited or ground) should then be multiplied by the corresponding feed down fraction and summed appropriately in order to obtain the inclusive suppression factor. 

\subsection{Dynamical Models for Quarkonium Production}

Phenomenological modeling of quarkonium suppression in the QGP is complicated by the fact that the matter generated in a heavy ion collision only exists in the plasma state for a few fm/c and during this very short time the properties of the QGP, such as the local temperature, rapidly evolve.  At the LHC energies, for example, in central collisions the QGP is formed at $\tau = \tau_0 \simeq 0.2$ fm/c after the initial nuclear impact and has a temperature on the order of $600~$MeV; however, the plasma rapidly cools to a freeze-out temperature $T  = T_f \simeq 150$ MeV within approximately $7~$fm/c. For non-central collisions, the initial temperature decreases and, as a consequence, the lifetime of the plasma becomes shorter.  In addition, the matter generated at all centralities has (on average) a spatial profile with the highest temperatures in the center of the overlap region and lower temperatures towards the edges.  In order to make quantitatively reliable predictions for quarkonium suppression, it is necessary to use dynamical models which can accurately describe the full spatio-temporal evolution of the QGP.

\begin{table}[t]
\tbl{Feed down fractions for (left) $J/\psi$  and (right) $\Upsilon$ production.  Quoted numbers internal to the charmonium system are world averages presented in Ref.~\protect\refcite{Faccioli:2008ir}.  The $J/\psi$ bottom hadron ($b$-hadron) feed down fraction is estimated from Ref.~\protect\refcite{Acosta:2004yw}.  The direct $J/\psi$ production fraction is estimated by requiring that the sum of the listed feed down fractions equal one.  Bottomonium feed down fractions are quoted from Ref.~\protect\refcite{Affolder:1999wm}.
}
{
\begin{tabular*}{.4\linewidth}{@{\extracolsep{\fill}}|l|l|}
\hline
\multicolumn{2}{|c|}{$J/\psi$ feed down fractions}\\\hline
{\bf Mechanism} & {\bf \% $\pm$ Error} \\\hline
Direct Production & 41 $\pm$ 17 \\\hline
$\Psi(2s)$ decay&  8.1 $\pm$ 0.3$\;$\cite{Faccioli:2008ir} \\\hline
$\chi_{c1}$ decay&  25 $\pm$ 5.0$\;$\cite{Faccioli:2008ir} \\\hline
$b$-hadron decay& 8.1 $\pm$ 3.2$\;$\cite{Acosta:2004yw} \\\hline
\multicolumn{2}{c}{}
\end{tabular*}
\hspace{4mm}
\begin{tabular*}{.5\linewidth}{@{\extracolsep{\fill}}|l|l|}
\hline
\multicolumn{2}{|c|}{$\Upsilon(1s)$ feed down fractions}\\\hline
{\bf Mechanism} & {\bf \% $\pm$ Stat $\pm$ Sys}$\;$\cite{Affolder:1999wm} \\\hline
Direct Production & 50.9 $\pm$ 8.2 $\pm$ 9.0 \\\hline
$\Upsilon(2s)$ decay&  10.7 $\pm$ 7.7 $\pm$ 4.8 \\\hline
$\Upsilon(3s)$ decay&  0.8 $\pm$ 0.6 $\pm$ 0.4 \\\hline
$\chi_{b1}$ decay&  27.1 $\pm$ 6.9 $\pm$ 4.4 \\\hline
$\chi_{b2}$ decay&  10.5 $\pm$ 4.4 $\pm$ 1.4 \\\hline
\end{tabular*}
}
\label{tab:feeddown}
\end{table}

Different groups have employed a variety of different models including basic ``fireball'' models (see e.g. Refs.~\refcite{Song:2010er,Emerick:2011xu}), ideal (2+1)-dimensional hydrodynamics including fluctuations (see e.g. Refs.~\refcite{Song:2011xi,Song:2011nu,Song:2011ev}), and (resummed) anisotropic viscous hydrodynamics\cite{Strickland:2011mw,Strickland:2011aa,Strickland:2012cq}.  At the very least, these dynamical models provide information concerning the spatio-temporal evolution of the local QGP energy density; however, they can additionally provide information about, for example, the local flow velocity of the medium, local flow velocity of the heavy quark states themselves, the degree of momentum-space isotropy of the QGP, etc.  This dynamical information must be integrated in proper-time in order arrive at the final predictions for the level of suppression for each quarkonium state under consideration.

There are also other models that try to explain quarkonium production and  suppression in heavy ion collisions. Most of these models ignore possible plasma anisotropies and assume complete local thermal equilibrium. Some of the models currently available are based on statistical recombination \cite{Andronic:2006ky}; statistical recombination and dissociation rates \cite{Zhao:2007hh,Zhao:2011cv,Emerick:2011xu} (this is the so-called two component model); or sequential melting \cite{Karsch:2005nk}. For loosely bound quarkonium states, as well as for unbound $Q \overline Q~$, a more realistic model has been developed \cite{Young:2008he}. This incorporates a realistic modeling of the expanding hot medium and a realistic potential, and it is relevant, in particular, for the understanding of charmonium production in heavy ion collisions since, as we have seen in the previous section, no charmonium states can exist deep in the deconfined state.

The bulk evolution of the matter produced in heavy-ion collisions is reasonably well-described by hydrodynamics (see e.g. Ref.~\refcite{Teaney:2009qa} for a recent review). The large heavy quark mass makes it possible to model its interaction with the medium by Langevin dynamics \cite{Moore:2004tg}. Such an approach successfully  describes the anisotropic flow of charm quarks observed at RHIC \cite{Moore:2004tg,vanHees:2004gq} (see also the review in Ref.~\refcite{Rapp:2009my} and references therein).  Potential models have shown that, in the absence of bound states, the $Q\overline Q$ pairs are correlated in space \cite{Mocsy:2007yj,Mocsy:2007jz}. This correlation can be modeled classically using Langevin dynamics, including a drag force and a random force between the $Q$ (or $\overline Q$) and the medium, as well as the forces between the $Q$ and $\overline Q$ described by the potential. It was shown that a model combining an ideal hydrodynamic expansion of the medium with a description of the correlated $Q \overline Q$ pair dynamics by the Langevin equation can describe charmonium suppression at RHIC quite well \cite{Young:2008he}. In particular, this model can explain why, despite the fact that a deconfined medium is created at RHIC, there is only a $40-50\%$ suppression in the charmonium yield. The attractive potential and the finite lifetime of the system prevents the complete de-correlation of some of the $Q \overline Q$ pairs \cite{Young:2008he}. Once the matter has cooled sufficiently, these residual correlations make it possible for the $Q$ and $\overline Q$ to form a bound state. The model was further developed to include a realistic viscous hydrodynamic description of the medium \cite{Young:2011ug}.  Production of charmonium states resulting from recombination of originally uncorrelated $c\bar c$ pairs, so-called off-diagonal recombination, can be also calculated in this framework \cite{Young:2009tj,Young:2011ug}.  The off-diagonal recombination is sub-dominant at RHIC \cite{Young:2009tj}, but is the dominant mechanism of $J/\psi$ production at the LHC \cite{Young:2011ug}. In particular, the calculation in Ref. \cite{Young:2011ug} was able to explain the very mild centrality dependence of the $J/\psi$ production at LHC.

\subsection{Cold Nuclear Matter Effects}

The goal of studying quarkonium suppression in nucleus-nucleus ($AB$) collisions is to use them as a probe to determine information about the QGP; however, in practice one has to be careful since propagation in the QGP is not the only way in which quarkonium yields can be affected.
As mentioned in the introduction, already in proton-nucleus ($pA$) collisions one sees modifications of quarkonium production. In order to compute the suppression of quarkonium states in a QGP one needs to know the initial spectrum of states including possible modifications from scattering, interference, absorption, and energy loss while the precursor states propagate through the nucleus itself.  Such modifications are lumped together into what are known as {\em cold nuclear matter} (CNM) {\em effects}.  

There are several CNM effects of importance.  The first to consider is modification of nuclear parton distribution functions (nPDFs).  Due to this effect, in different kinematical regions quarkonium production can be suppressed (shadowing) and enhanced (anti-shadowing) relative to proton-proton ($pp$) collisions (we will lump both together as ``shadowing'' in the remainder). The key inputs necessary are the quark and gluon nPDFs. The quark nPDFs are fairly well-constrained by nuclear deep-inelastic scattering (nDIS) experiments; however, the gluon distributions must be inferred from scaling violations and conservation laws.  

In order to compute CNM shadowing, typically parton distribution functions such as CTEQ6 \cite{Pumplin:2002vw} are used in conjunction with different shadowing parameterizations, e.g. EKS98 \cite{Eskola:1998iy,Eskola:1998df}, nDSg \cite{deFlorian:2003qf}, HKN \cite{Hirai:2004wq}, EPS08 \cite{Eskola:2008ca}, or EPS09 \cite{Eskola:2009uj}. Unfortunately, there is a sizable variation of the resulting estimates depending on the choice of shadowing parameterization and in some cases even within the uncertainties quoted by an individual parameterization, e.g. EPS09.  In a recent work, a study of CNM shadowing effects on $J/\psi$ and $\Upsilon$ production at the top LHC energy of $\sqrt{s_{NN}} = 7$ TeV was performed \cite{Vogt:2010aa}. The analysis found that for central Pb-Pb collisions at $y=0$, the nuclear suppression factor due to CNM shadowing effects alone was on the order of $R_{AA} \simeq 0.5 \pm 0.4$ for the $J/\psi$ and $R_{AA} \simeq 0.65 \pm 0.3$ for the $\Upsilon$, with the uncertainties estimated by varying the EPS09 parameters within their stated uncertainties. The central values indicate that there could indeed be sizable CNM shadowing effects on both the $J/\psi$ and $\Upsilon$ at LHC energies; however, more experimental and theoretical work is needed to further eliminate uncertainties in the nPDFs.  In this context we also point to Ref.~ \cite{Ferreiro:2011xy} in which an analysis of CNM effects on $\Upsilon$ production was performed, and found at central rapidity shadowing on the order of 10\% in p-Pb collisions at LHC energies.

Another CNM effect which must be taken into consideration is the nuclear absorption of quarkonium bound states after they form. In general the effective absorption cross section depends on the collision energy and the rapidity of the observed state.  At sufficiently high energies, one expects negligible absorption cross sections since the formation times for these states are higher than the amount of time they spend traversing the target.  Early works assumed a constant absorption cross section which might be different for singlet and octet states \cite{Povh:1987ju,Blaizot:1988ec,Blaizot:1988hh,Blaizot:1989de,Kharzeev:1995br,Vogt:1999dw}; however, more recent studies \cite{Vogt:2004dh,Lourenco:2008sk} show that the effective absorption cross section decreases with increasing energy which seems to be in agreement with recent experimental findings.~\cite{Scomparin:2009tg,Arnaldi:2009it}.  In addition, we once again point to Ref.~\cite{Ferreiro:2011xy} which found that for the $\Upsilon$ the survival probability is close to unity, implying a small effective absorption cross section for bottomonium states.  In this context, we mention that one also has to consider energy loss during the traversal of the target.  For a discussion of different energy loss models and their impact on quarkonium production we refer the reader to Ref.~\refcite{Vogt:1999dw}.

Finally, we mention that at very high energies the small-$x$ partonic wave function of a heavy ion is expected to be dominated by a highly-occupied saturated gluonic state called the color glass condensate (CGC) \cite{McLerran:2001sr}. The CGC is expected to play an important role in the initial state suppression of charmonium production, since the saturation scale $Q_s$ is on the same order of magnitude as the charm quark mass \cite{Kharzeev:2003sk}. This effect could be particularly important for forward $J/\psi$ production \cite{Kharzeev:2008nw}.  The effect is expected to be smaller for bottomonium states due to their larger mass; however, it should be taken into account nonetheless.  The $J/\psi$ nuclear modification factor was studied in Refs. \refcite{Dominguez:2011cy} and \refcite{Kharzeev:2012py} in the CGC framework, and it has been shown that $R_{dA}$ can be well-described in this approach \cite{Dominguez:2011cy}.


\section{Conclusions and Outlook}

In this paper we have attempted to make a compact review of (a) the theory of quarkonium states in the quark gluon plasma and (b) the phenomenology of quarkonium production in relativistic heavy ion collisions.  
There has been significant progress since the original predictions of Matsui and Satz.  
Thanks to first principles calculations available from lattice QCD it is possible to make non-perturbative studies of the behavior 
of the heavy quark potential and quarkonium spectral functions.  
Additionally, there has been significant progress made in recent years within the effective theory approach at zero and finite temperature 
allowing a firm basis for systematically improvable potential model calculations.  Using the effective theory approach, 
somewhat unanticipated imaginary-valued contributions to the quarkonium potential have been identified in recent years.  
These imaginary-valued contributions can be shown to be directly related to the in-medium decay of quarkonium states.  
Depending on the QGP energy density and state under consideration, the implied in-medium widths are on the order of 20-100 MeV, 
which is orders of magnitude higher than the corresponding vacuum widths.  

Such large widths imply in-medium heavy-quarkonium lifetimes on the order of fm/c, thereby causing a blurring of the heretofore assumed pseudo-sharp 
thresholds for the dissolution of different quarkonium states, commonly referred to as dissociation temperatures.  
As a result, it is necessary to include the effect of such imaginary-valued contributions to the potential in phenomenological models 
and a few authors have already begun to implement the necessary dynamical models.  In addition, phenomenological models of quarkonium suppression 
are seeing an across-the-board improvement in terms of the underlying dynamical models being used.  
Most practitioners are moving towards using some variant of hydrodynamics (ideal, viscous, or anisotropic) which allows the dynamics of 
collective flow and heavy quarkonium suppression to be described simultaneously.  In addition, advances have been made in the modeling of the 
process of recombination, even including effects of $Q\bar{Q}$ spatial correlations in a statistical approach utilizing Langevin dynamics.  
Needless to say, further work is necessary but in recent years many theoretical groups have been contributing to key advances in this area.

Looking forward, as the question of quarkonium suppression becomes more quantitative, it will become increasingly important to further elucidate 
cold nuclear matter effects on quarkonium production in heavy ion collisions.  The $pA$ runs at LHC will provide much-needed 
input which will hopefully help to constrain models of cold nuclear matter effects.  Additionally, further refinements of the space-time models 
of the quark gluon plasma are necessary including the incorporation of fluctuations in the viscous hydrodynamical treatments, 
further investigation of the impact of momentum-space anisotropies, effects of finite velocity of quarkonium states relative to the medium, etc.  
Through the concerted effort of both theoretical and experimental collaborations, 
we hope to see continued progress towards a better understanding of the production and dynamics of quarkonium in the quark gluon plasma.


\section*{Acknowledgments}

The work of P.P. was supported by U.S. Department of Energy under Contract No. DE-AC02-98CH10886.   
M.S. was supported by NSF grant No.~PHY-1068765.

\bibliographystyle{iopart-num}
\bibliography{qonium}

\end{document}